\documentclass{article}
\usepackage{textcomp}
\usepackage[utf8]{inputenc}
\usepackage{float}
\usepackage{amsmath}
\usepackage{amssymb}
\usepackage{graphicx}
\usepackage{geometry}
\usepackage[bookmarks=false,
 breaklinks=false,pdfborder={0 0 1},backref=section,colorlinks=false]
 {hyperref}
\hypersetup{
 hidelinks}

\makeatletter

\usepackage{amsfonts}

\usepackage[section]{placeins}

\usepackage{algorithmic}

\usepackage{xcolor}

\usepackage{cite}

\graphicspath{{figures/}}
\usepackage{caption}
\usepackage{subcaption}

\makeatother

\begin{document}
\global\long\def\abstractname{Abstract}%
\global\long\def\figurename{Figure}%
\global\long\def\tablename{Table}%
\global\long\def\refname{References}%

\title{A General-purpose Solver of Fourier Neural Swarm Operator Towards Accurate and Efficient Mechanical Modeling of Ultra Large Composite Materials}

\author{Lekun Gao$^{1}$, Shaohua Chen$^{2,*}$}

\date{}
\maketitle

\begin{center}
	$^{1}$ College of Astronautics, Nanjing University of Aeronautics
	and Astronautics, Yudao Street 29, Nanjing, China\\
	$^{2}$ College of Energy and Power Engineering, Nanjing University
	of Aeronautics and Astronautics, Yudao Street 29, Nanjing, China
\end{center}

\begin{abstract}
Composite media with complex microstructures exhibit highly tailorable mechanical properties but remain challenging to model efficiently and accurately. Conventional homogenization often oversimplifies microstructural effects, whereas multiscale approaches typically require costly coupling across spatial and temporal scales. To address these limitations, we propose a two-scale neural-swarm framework for large-scale mechanical modeling of heterogeneous composites. At the local scale, the mechanical characteristics of representative microstructural features are encoded into building-block Fourier neural operators (FNOs) using level-set representations. At the global scale, these pretrained FNOs are assembled into an FNO swarm according to the spatial distribution of microstructural constituents. A coarse-mesh finite element model is employed to provide global physical guidance, while Schwarz iteration is used to synchronize neighboring FNOs and enforce consistency across shared interfaces. The proposed framework is validated through nonlinear simulations of SiC–Al composites with diverse microstructural configurations. Compared with nonlinear finite element analysis, the FNO-swarm method achieves comparable accuracy while reducing computational cost by orders of magnitude. For an extreme dual-property SiC–Al composite containing more than a billion nodal points, the proposed approach predicts the mechanical response within approximately one hour, demonstrating exceptional scalability. Furthermore, the framework naturally accommodates arbitrary Dirichlet boundary conditions and complex domain geometries. The proposed neural-swarm strategy provides a robust and scalable paradigm for large-scale mechanics simulations, reconciling the longstanding trade-off between computational efficiency and physical fidelity in heterogeneous materials modeling.
\end{abstract}
\vspace{1em}
 Keywords: Artificial neural network, Fourier neural operator, Composite
material, Swarm Intelligence, Finite element method

\section{Introduction}

Due to their flexible and designable microstructural characteristics,
complex-structured composites have being increasingly studied and
employed\cite{2} . Accurately and efficiently predicting their mechanical
responses under different working conditions has become an increasingly
urgent demand. However, these materials exhibit distinct hierarchical
features\cite{EVANS1994177,1} , and their macroscopic mechanical
properties are highly dependent on constituents distribution, interface
evolution, and damage mechanisms at the meso- and microscopic scales.
How to accurately simulate their multiscale mechanical behaviors has
always posed a tremendous challenge\cite{Fish2021MesoscopicAM}.

Traditional numerical methods in the field of multiscale mechanics
are deeply mired in the contradiction between computational efficiency
and solution accuracy. Microscopic simulation techniques such as molecular
dynamics (MD) are constrained by extremely small time steps and spatial
scales, making it difficult to apply them directly to predicting macroscopic
engineering component responses \cite{Fish2021MesoscopicAM}. The
Phase field method, when resolving mesoscale diffuse interfaces, relies
on extremely high local mesh resolutions, and the resulting massive
degrees of freedom leave it facing unacceptable computational costs
in complex evolution calculations \cite{WU20201}. As the mainstream
simulation tool for solid mechanics, the finite element method (FEM),
when handling multiphase media such as large-particle inclusions,
must generate highly refined meshes locally to accurately capture
intense local stress concentrations and nonlinear evolution effects
at phase interfaces. This leads to a sharp explosion in the dimensionality
of the global system stiffness matrix, drastically escalating computational
and memory overhead, thereby limiting its application in large-scale
multiscale simulations\cite{10.1016/j.cam.2009.08.077}.

To ease the computational cost, researchers have developed a variety
of multiscale methods. The first category is hierarchical multiscale
methods, whose core idea is ``bottom-up.” Computations are performed
independently at different scales, and lower-scale computational results
are passed to higher scales as equivalent parameters. An typical example
is the concurrent method introduced by Rodney Hill and other scholars\cite{Hill1963Elastic,Hill1965Self},
which derives equivalent macroscopic effective properties by analyzing
microstructural properties. This significantly compresses macroscopic
computational degrees of freedom and has achieved success in predicting
the global average response of materials. Building on this, Gamra
et al. developed a decoupled second-order gradient homogenization
method \cite{GAMRA2026120070}, which can effectively handle highly
nonlinear large deformation problems with extreme stiffness contrast
between the matrix and inclusions. Nevertheless, during scaling up,
critical local mechanical information such as stress concentrations
might get lost\cite{Feyel_Chaboche_2000}. How to retain microscopic
details while predicting macroscopic mechanical behavior of materials
under limited computational cost remains an unresolved problem for
multiscale method.

The second category is domain decomposition, namely synchronously
resolving macroscopic and microscopic models in different spatial
regions. A representative approach is the over-lapping Schwarz method
introduced by Pierre-Louis Lions \cite{Lions1988SchwarzI,Lions1989SchwarzII,Lions1990SchwarzIII}.
This method reduces the dimensionality of the original systems into
multiple local subproblems, achieving acceleration in global solving.
Subsequently, Liu et al. improved the overlapping Schwarz method for
nonlinear systems \cite{Liu2024Overlapping}, rigorously proving that
a system preconditioned by nonlinear multiplicative Schwarz has the
exact same unique solution as the original global nonlinear physical
system. However, domain decomposition still faces various issues,
such as the difficulty in balancing kinematic and mechanical coordination
during domain stitching, and spurious reflections of high-frequency
stress waves at interfaces inducing strong numerical noise, making
it difficult to predict the true local stress field.

Recently, the booming development of machine learning (ML) technology
has brought new hope to solve the aforementioned problems in two routes\cite{JAIN2024101189,Liu2019,KIRCHDOERFER201681,Bessa2017}.
The first one is the development of deep learning-based PDE solvers.
A typical example if the physics-informed neural networks (PINNs)
proposed by Raissi et al\cite{Raissi2019}.This method can embed physical
constraints, such as the constitutive relations of composite materials,
directly into the neural network’s loss function, thereby freeing
it from a heavy reliance on large-scale high-fidelity data labels\cite{Haghighat2021}.
Subsequently, drawing on the local basis function construction of
traditional FEM, Ben Moseley et al. proposed finite basis physics-informed
neural networks (FBPINNs) \cite{Moseley2021FiniteBP}, achieving high-accuracy
parallel scaling of models across scales. Based on classical laminated
plate theory, Sahar et al. utilized PINNs to develop a model specifically
for accurately predicting the bending deformation behavior of laminated
composite plates\cite{WANG2025113014}. This model requires only a
small amount of strain data from damaged zones to inversely determine
accurate damage evolution parameters, breaking the bottleneck of physical
consistency in purely data-driven models.

The second route is the so-called operator learning\cite{JMLR:v24:21-1524},
represented by the milestone Fourier neural operator (FNO) proposed
by Li et al\cite{li2021fourier}. This fundamentally broke through
the bottleneck of traditional neural networks being restricted to
finite-dimensional vector space mappings, achieving mapping learning
between infinite-dimensional function spaces. Since then, advanced
architectures such as deep operator networks (DeepONet)\cite{2019DeepONet}
and physics-informed neural operators\cite{10.1145/3648506} (PINO)
have emerged successively. In this direction, Binh Huy Nguyen et al.
integrated FNO with an FFT-based computational micromechanics solver,
proving that FNO can extremely rapidly process microscopic unit cell
problems with arbitrary stiffness distributions and extreme material
contrasts \cite{NGUYEN2026106418}. Goswami et al. proposed the variational
DeepONet\cite{Goswami2021APV}, directly using energy functionals
from fracture mechanics as the loss function. It can predict damage
evolution and crack propagation paths in quasi-brittle materials under
arbitrary initial cracks and macroscopic loads. Furthermore, Wu et
al. introduced the physics-pretrained neural operator (PPNO)\cite{WU2026118799},
demonstrating excellent accuracy and scalability in the statistically
averaged random grain single-degree-of-freedom solution domain.

Although ML has demonstrated immense potential in solid mechanics,
it still faces two major issues compared with mature traditional numerical
methods. First, ML-based method usually has scalability issue. For
a single neural network model, the problem size is often hard-coded
into the input and output layer dimensions. Once the computational
domain changes size, the prediction result may deteriorate or even
become unusable. Moreover, some ML-based model requires that the computational
domain be regular shaped. For example, Liu et al. pointed out that
FNO fundamentally relies on the fast fourier transform (FFT), which
is mathematically restricted to rectangular and uniformly discretized
computational domains. However, real engineering structures are rarely
perfectly rectangular. As a result, standard FNO frameworks cannot
be directly applied to the vast majority of realistic physical domains
\cite{2023Fourier,NEURIPS2023_940a7634}. Therefore, a general-purpose
method has to solve the scalability issue, so that the ML-based model
can effectively deal with arbitrary problem size of computational
domain without constructing and training a new model.

Second, ML-base method needs to guarantee a decent convergence. Unlike
traditional numerical algorithms such as the Galerkin method, neither
physics-informed neural networks (PINNs) nor operator learning can
mathematically provide rigorous convergence guarantees for non-convex
problem, especially for large-scale problems. For PINNs, multiple
failure modes have been reported \cite{NEURIPS2021_df438e52}. M.
Z. Naser pointed out that PINN optimization is highly challenging,
and the network may converge to mathematically admissible yet physically
unrealistic states. Furthermore, PINNs trained under linear elastic
conditions may perform well when validating linear features, while
completely failing to capture nonlinear material failure phenomena
\cite{NASER2026111704}. From the perspective of neural tangent kernels
(NTK), Wang et al. theoretically analyzed the intrinsic limitations
of PINNs and demonstrated that, due to spectral bias, PINNs exhibit
severe over-smoothing when handling nonlinear problems, thereby obscuring
high-frequency physical features such as stress concentrations, turbulence,
and phase transitions \cite{WANG2022110768}. Regarding neural operators,
Zhang et al. systematically investigated the mathematical deficiencies
of latent-space neural operators. They explicitly noted that existing
latent-space updates in neural operators are essentially unconstrained
Euclidean space operations, leading to the so-called “latent drift”
phenomenon. In strongly nonlinear or high-frequency problems, small
errors can accumulate and amplify rapidly during inter-layer iterations
of the operator network, ultimately resulting in predictions that
violate physical consistency \cite{Zhang2026GeometricNO}.

Targeting these two major issues, we propose a two-scale FNO swarm
methodology with guaranteed scalability and convergence. First, we
employ the level set method to quantify the microstructural features
and use FEM to collect the mechanical behaviors of the simulation
domain. Then, we build and train FNO models to learn from the mechanical
behaviors corresponding to typical microstructural features. Next,
we integrate the FNOs into FNO swarm according to the microstructural
configuration of the simulation domain of interest. Finally, we utilize
Schwarz iteration to drive the collective inference of the FNOs to
reach systematic synchronization, whose output will be the desired
solution. In the remaining text, section 2 introduces the construction
of local building-block FNOs and global FNO swarm for SiC-Al composite
system; section 3 presents extensive numerical experiments to demonstrate
the accuracy and effectiveness of the FNO-swarm method; section 4
inspects the convergence, computational efficiency and robustness
of the FNO-swarm method. In section 4, we discuss the limitations
and future development directions of our method. In section 5, we
concludes this work with a few remarks.

\section{Construction of FNO and FNO swarm}

\subsection{SiC particle reinforced Al composite}

Particle-reinforced metal composites are ubiquitous in aerospace structures,
automotive components, energy equipment, etc., owing to their superior
specific strength, stiffness, thermal stability, and tailorability
compared with conventional monolithic metals. Among them, silicon
carbide (SiC) reinforced aluminum (Al) composite is widely studied
because they combine the lightweight and ductile characteristics of
aluminum with the high stiffness, strength, and wear resistance of
silicon carbide, resulting in excellent specific mechanical properties.
This study adopts the plane stress assumption for SiC-Al composite
system. The in-plane stress and strain are represented by the column
vectors $\sigma=[\sigma_{xx},\sigma_{yy},\sigma_{xy}]^{T}$ and $\varepsilon=[\varepsilon_{xx},\varepsilon_{yy},\gamma_{xy}]$,
respectively, where $\gamma=2\varepsilon_{xy}$ is the engineering
shear strain. Given a SiC-Al composite $\Omega$, the displacement
field $u$ is governed by following elliptic differential equation,
\[
-\nabla\cdot(\mathbb{D}(\mathbf{x};\epsilon(u)):\nabla_{s}u)=0\quad in\ \Omega
\]
where $\nabla_{s}u=\frac{1}{2}[\nabla\mathbf{u}+(\nabla\mathbf{u})^{T}]$
is the strain tensor, $\epsilon(u)$ is the local strain, and $\mathbb{D}(\mathbf{x},\epsilon(u))$
is the strain dependent stiffness tensor. For Al-SiC composite, 
\[
\mathbb{D}(\mathbf{x};\epsilon(u))=\begin{cases}
\mathbb{D}_{SiC}(\sigma_{vm}(\epsilon(u))) & \mathbf{x}\in\Omega_{SiC}\\
\mathbb{D}_{Al}(\sigma_{vm}(\epsilon(u))) & \mathbf{x}\in\Omega_{Al}
\end{cases}
\]
where $\mathbb{D}_{SiC}$ and $\mathbb{D}_{Al}$ are the intrinsic
stiffness tensor of SiC and Al, respectively. The SiC reinforcement
phase exhibit typical brittle characteristics, and its strength is
vastly higher than that of the aluminum alloy matrix. So when the
composite material undergoes significant plastic flow, the stress
level within the SiC particles usually remains far below their strength
limit. Therefore, we reasonably set $\mathbb{D}_{SiC}$ to be a constant.
The Al matrix is the main carrier of plastic dissipation in the composite.
Here we assume that in the small strain stage ($\leq0.3\%$), the
matrix remains linearly elastic; in the large deformation stage ($0.3\%\sim3\%$),
a $J_{2}$ invariant-based Ramberg-Osgood model is adopted to characterize
its strain hardening characteristics without a distinct yield
plateau\cite{1943Description,Tvergaard1990AnalysisOT}.
The elastic modulus $E_{Al}$ of the matrix can be explicitly expressed
as a function of the current von Mises equivalent stress $\sigma_{vm}$,
\[
E_{Al}(\sigma_{vm})=\frac{\sigma_{vm}}{\epsilon_{vm}}=\frac{E_{0}}{1+\alpha n(\frac{\sigma_{vm}}{\sigma_{0}})^{n-1}}
\]
where $\epsilon_{vm}$ is the von Mises equivalent strain with $\ensuremath{\epsilon_{vm}=\sqrt{\epsilon_{xx}^{2}+\epsilon_{yy}^{2}-\epsilon_{xx}\epsilon_{yy}+3\epsilon_{xy}^{2}}}$.
$\sigma_{vm}$ is von Mises stress with $\ensuremath{\sigma_{vm}=\sqrt{\sigma_{xx}^{2}+\sigma_{yy}^{2}-\sigma_{xx}\sigma_{yy}+3\sigma_{xy}^{2}}}$
,$E_{0}$ is the initial elastic modulus of the matrix, $\alpha$
is a dimensionless material constant , $n$ is the strain hardening
exponent, and $\sigma_{0}$ is the reference yield stress. In this
study, $E_{0}=70\text{ GPa}$, $\alpha=0.5$, $n=7$, and $\sigma_{0}=0.25\text{ GPa}$.
\begin{figure}[H]
\centering \includegraphics[width=0.95\textwidth]{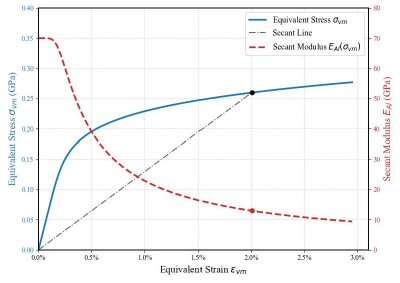} \caption{Nonlinear constitutive response of the aluminum (Al) matrix under
the Ramberg--Osgood model based on the $J_{2}$ invariant. The blue
solid curve on the left vertical axis represents the evolution of
the von Mises equivalent stress $\sigma_{vm}$ with respect to the
equivalent strain $\epsilon_{vm}$. The red dashed curve on the right
vertical axis shows the corresponding nonlinear degradation of the
secant modulus $E_{Al}(\sigma_{vm})$. The dash-dotted line illustrates,
as an example, the state at $\sigma_{vm}=0.26Gpa$ (corresponding
to an equivalent strain of approximately 2.0\%).}
\end{figure}

\subsection{Composite microstructure quantification}

Since FNO model is based on Fourier transform, it inherently has limitations.
For example, when dealing with irregular computational domains or
complex internal structures, Fourier transform often exhibits the
Gibbs phenomenon near geometric discontinuities, leading to significant
accuracy degradation in capturing critical regions such as stress
concentrations. Since Fourier transform requires periodicity, it may
lead to spectral leakage near non-periodic boundaries. Moreover, in
practical usage, FNO focuses on a few low-frequency modes while omitting
the high-frequency ones. Such truncation operation usually introduces
spectral bias and results in information loss of local sharp features.
Considering the discontinuity of the composite microstructure, how
to mathematically smoothen the interphase boundary is critical for
the accuracy and efficiency of FNO modeling.

Level set method (LSM) is an approach typically used to track interface
and shape evolution in the field of solid-liquid coupling, topology
optimization, etc. \cite{345,2017A,Allaire2004StructuralOU}. Its
core idea is to represent a low-dimensional boundary as the zero-level
set of a higher-dimensional function \cite{1988Fronts}. Here, we
use it to convert the originally discrete SiC-Al microstructure into
smooth functional space. Specifically, let the overall computational
domain of the composite material be $\Omega\subset R^{2}$, which
contains the reinforcement phase $\Omega_{inclusion}$ and the matrix
phase $\Omega_{matrix}$, and the interface between the two denoted
by $\Gamma_{interface}$. On this basis, a signed distance function
(SDF) $\phi(\mathbf{\mathbf{x}}):\Omega\rightarrow\mathbb{R}$ is
defined as follows: 
\[
\phi(\mathbf{x})=sgn(\mathbf{x})min||\mathbf{x}-\mathbf{y}||_{2},\mathbf{y}\in\Gamma_{interface}
\]
where 
\[
sgn(\mathbf{x})=\begin{cases}
1 & \mathbf{x}\in\Omega_{inclusion}\\
0 & \mathbf{x}\in\varGamma_{interface}\\
-1 & \mathbf{x}\in\Omega_{matrix}
\end{cases}
\]
As a result, $\phi(\mathbf{x})$ ca be used as a smooth characterization
of the composite microstructure, i.e., 
\[
\phi(\mathbf{x})\begin{cases}
>0 & \mathbf{x}\in\Omega_{inclusion}\\
=0 & \mathbf{x}\in\varGamma_{interface}\\
<0 & \mathbf{x}\in\Omega_{matrix}
\end{cases}
\]

Since the gradient norm of $\phi(\mathbf{x})$ is constantly equal
to 1, namely, $|\nabla\phi(\mathbf{x})|=1$, it possesses excellent
differential geometric property across the entire domain $\Omega$,
which is of great convenience in training FNOs. 
\begin{figure}[H]
\centering \includegraphics[width=0.9\textwidth]{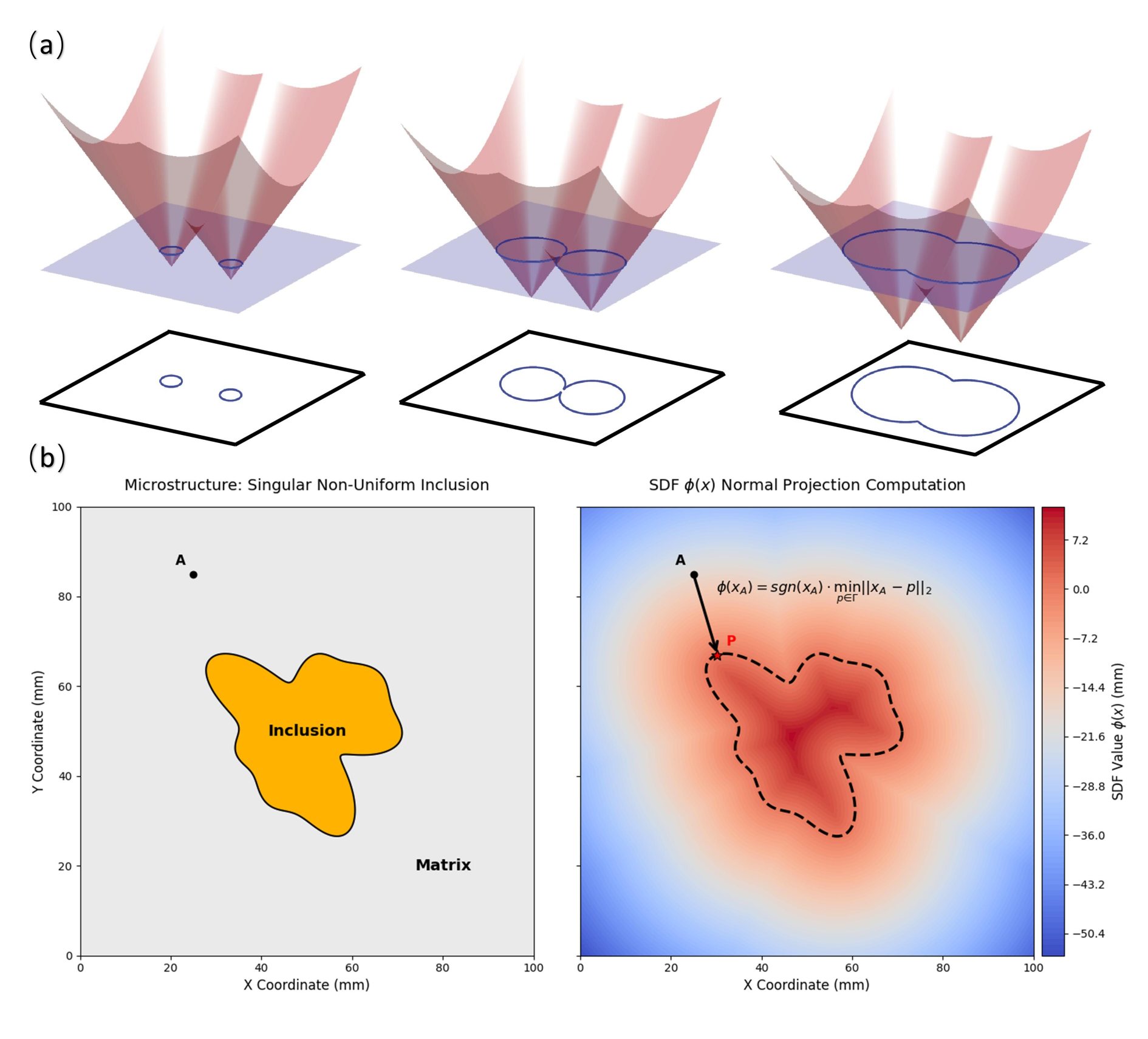}
\caption{(a)Illustration of using LSM to compresses three-dimensional features
into two dimensions\cite{2017A}. (b) an example of using LSM to convert
discontinuous composite microstructure into continuous signed distance
field (SDF).}
\label{fig:level_set} 
\end{figure}

\subsection{Collection of mechanical responses}

Before introducing the FNO-swarm method, we assume that the composite
material of interest consists of finite number of typical microstructure
features. Then, we can construct a FNO model for each microstructure
feature. Considering that the FNO model to be constructed relies on
the fast Fourier transform (FFT), which requires input and output
data be defined on a uniform grid, we draw a series of equidistance
points on the square shaped domain of typical microstructure. Specifically,
We generate a regular $N\times N$ Cartesian grids within the domain
and perform Delaunay triangulation on these points. The material property
distribution (SiC reinforcement and Al matrix) based on the signed
distance function $\phi(\mathbf{x})$ is then explicitly assigned
to these uniformly distributed triangular elements. Fig.3 illustrates
how the discretization process is carried out.

\begin{figure}[H]
\centering \includegraphics[width=0.95\textwidth]{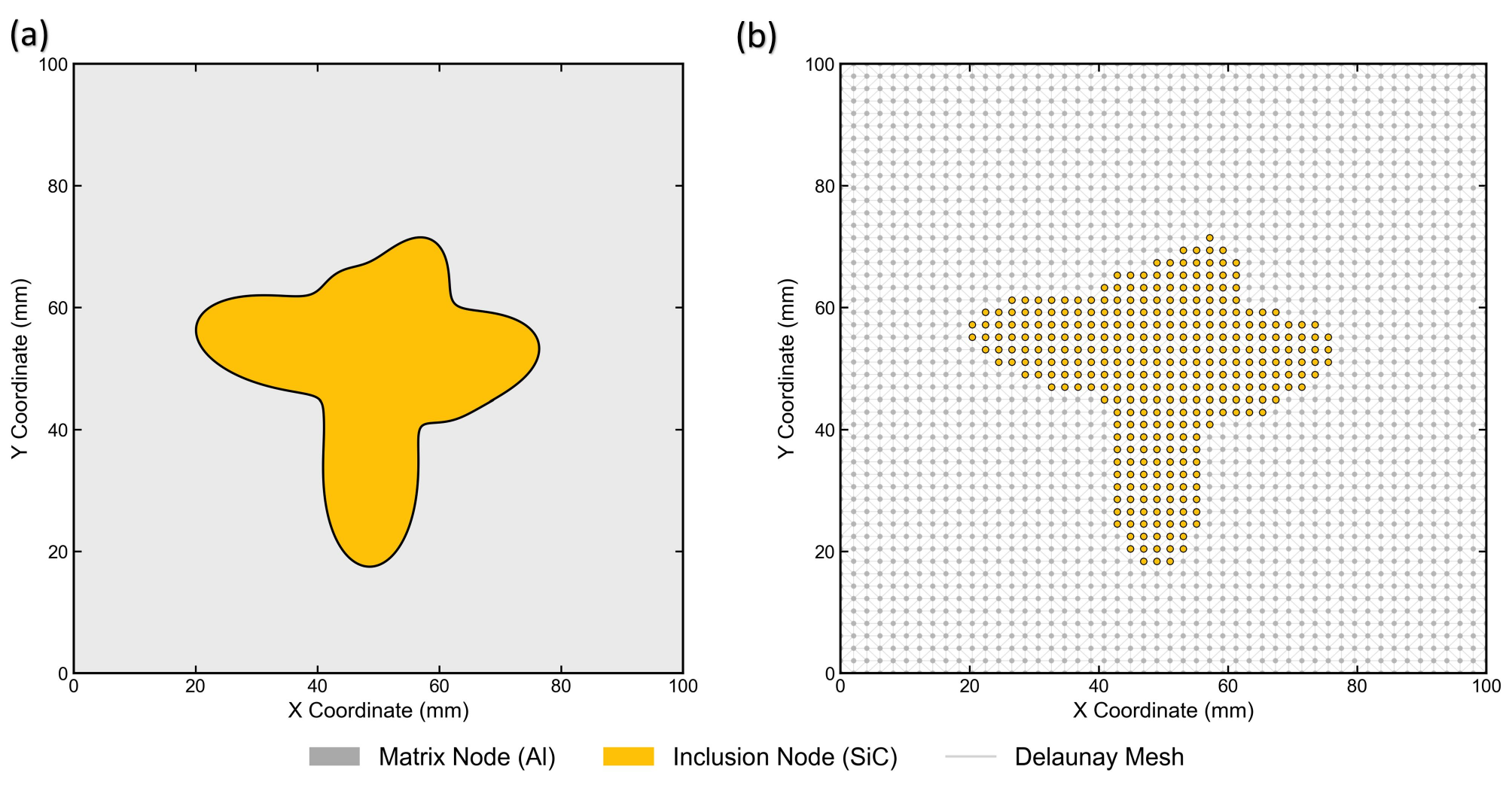}
\caption{(a) schematic illustration of typical microstructure feature of SiC-Al
composite, where the gray region represents the Al matrix and the
yellow region of the reinforcement SiC particle, (b)the structured mesh discretization
obtained by Delaunay triangulation.}
\end{figure}

Given the meshed domain, we inspect its mechanical behaviors by randomly
enforcing Dirichlet boundary conditions (BCs) on the exterior edges
and calculate the resulted mechanical responses. Specifically, the
displacement component along each edge consists of a linear baseline
superimposed by a nonlinear perturbation of trigonometric or Gaussian
distribution function, as shown in Fig.4. We employ FEM with under-relaxed
Picard iteration method to determine the displacements on the interior
nodes of the microstructural dmain by solving $\mathbf{KU=F}$, with
$\mathbf{K}$ being the global stiffness matrix, $\mathbf{U}$ being
the global displacement field, and $\mathbf{F}$ being the global
nodal force vector. When enforcing the Dirichlet BCs, the penalty
method is employed by introducing extremely large penalty coefficients
to the diagonal elements of the $\mathbf{K}$ and the corresponding
nodal load vector $\mathbf{F}$. In the solution phase, a sparse solver
based on LU decomposition is utilized to obtain the global vector
of nodal displacement. The computational effort and accuracy of such
FEM calculation is closely related to the number of $N$. For training
data collection, we set $N=50$ to guarantee the physical fidelity
of the calculation result. While for initialization of the FNO swarm,
$N$ can take very small value as we will see later.

\begin{figure}[H]
\centering \includegraphics[width=0.95\textwidth]{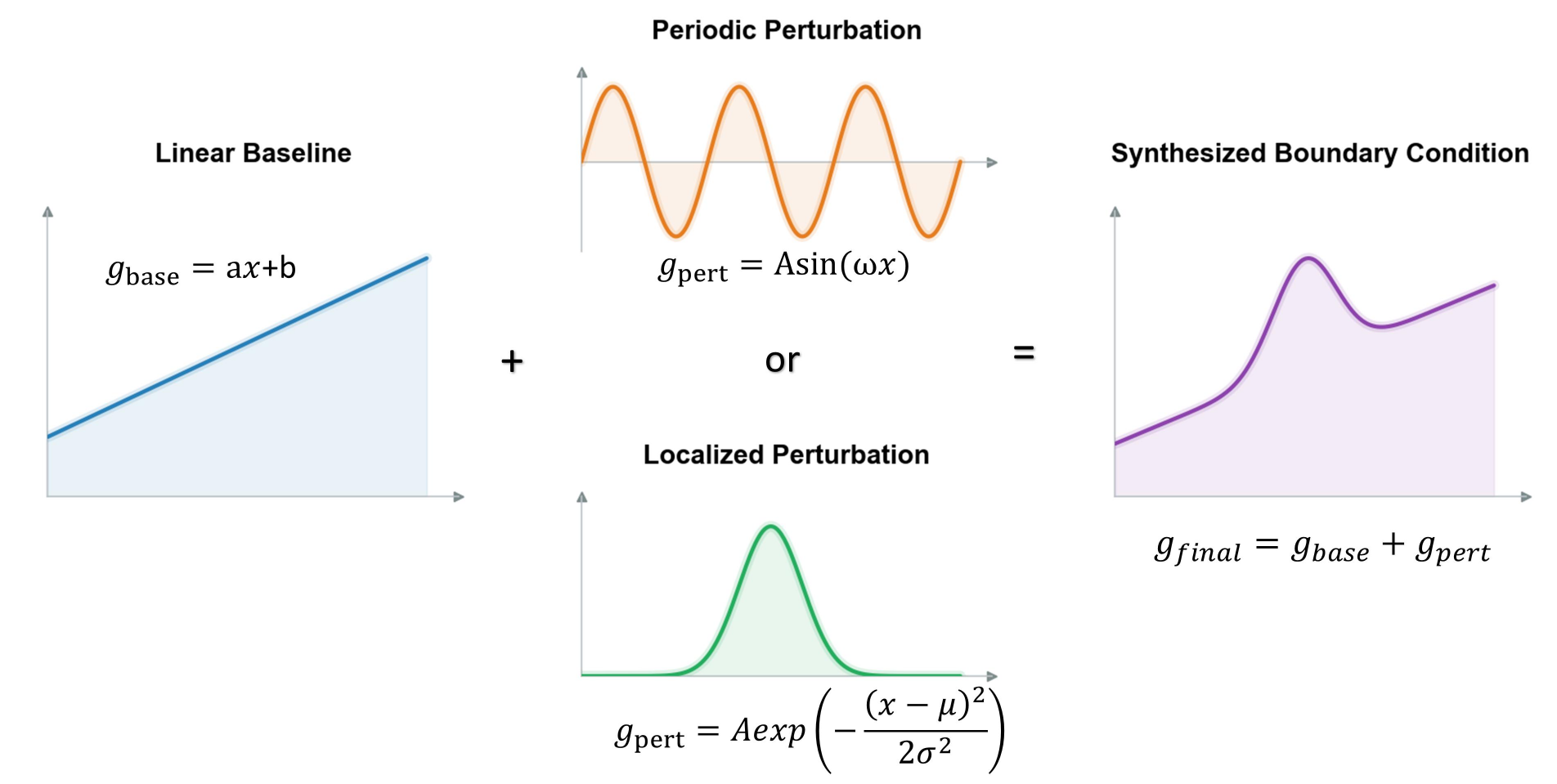}
\caption{Schematic illustration of Dirichlet BC generation. The Dirichlet BC
along each edge consists of two parts, $g_{base}$ and $g_{pert}$.
$g_{base}$ represents the base term which is a linear distribution,
while $g_{pert}$ is an extra nonlinear term, which can take the form
of trigonometric or Gaussian function.}
\end{figure}

\subsection{Construction of FNOs}

Let $\Omega$ be a bounded open set ($\Omega\subset\mathbb{R}^{d}$),on
which two Banach function spaces are defined, i.e., the input space
$\mathcal{A}=\mathcal{A}(\Omega;\mathbb{R}^{d_{in}})$ and the output
space $\mathcal{U}=\mathcal{U}(\Omega;\mathbb{R}^{d_{out}})$, composed
of square-integrable functions taking values in $\mathbb{R}^{d_{in}},\mathbb{R}^{d_{out}}$,
respectively. Our objective is to construct an FNO to approximate
the nonlinear operator $\mathcal{G}:\mathcal{A}\longrightarrow\mathcal{U}$,
such that for any input function $a\in\mathcal{A}$, it can output
the corresponding $u\in\mathcal{U}$. The overall FNO architecture
consists primarily of three parts, the lifting layer, the Fourier
layers, and the projection layer. The lifting layer maps the input
function $a(x)$ into to a high-dimensional feature space via a fully
connected neural network $P$, yielding the initial hidden state $v_{0}=P(a(x))$.
The stacked Fourier layers updates the hidden state $v_{l}$ .The
update in each Fourier layer includes a global integral operator and
a local linear transformation. The projection layer maps the output
$v_{L}(x)$ of the final Fourier layer back to the target output dimension
via another fully connected neural network $Q$, yielding the final
prediction $u(x)=Q(v_{L}(x))$. Herein, the evolution rule of the
$l$-th Fourier layer is defined as: 
\[
v_{l+1}(x)=\sigma(Wv_{l}(x)+(\mathcal{K}v_{l})(x))
\]
where $\sigma$ is the activation function, $W$ is the linear transformation
matrix acting on the channel dimension, and $\mathcal{K}$ is the
kernel integral operator, $(\mathcal{K}v_{l})(x)=\int_{\Omega}k(x,y)v_{l}(y)dy$
with $k$ being a continuous kernel function. The kernel integral
operator essentially captures global interactions within the domain
via Green’s function. However, directly calculating the above global
integral has a computational complexity of $O(N^{2})$. Therefore,
assuming the kernel is translation-invariant $k(x,y)=k(x-y)$, convolution
in the spatial domain is equivalent to multiplication in the frequency
domain. Thus, in FNO, the operator $\mathcal{K}$ is parameterized
as: 
\[
(\mathcal{K}v_{l})(x)=\mathcal{F}^{-1}(R_{l}\cdot\mathcal{F}(v_{l}))(x)
\]
$\mathcal{F}$ and $\mathcal{F}^{-1}$ are the fast Fourier transform
and its inverse transform, $R_{l}$ is a complex-valued weight tensor
learned directly in the frequency domain. Through low-pass truncation,
$R_{l}$ achieves global information interaction and feature extraction
with quasi-linear complexity $O(NlogN)$. 
\begin{figure}[H]
\centering \includegraphics[width=0.9\textwidth]{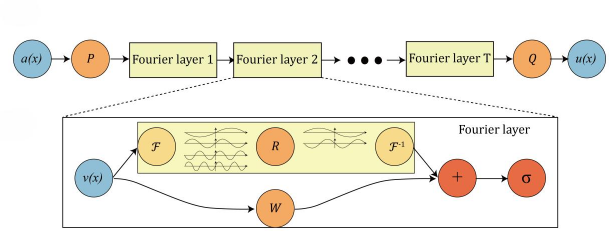} \caption{Schematic of the building-block FNO architecture\cite{li2021fourier} }
\label{fig:fno} 
\end{figure}

Let us now construct FNO for SiC-Al composite. Considering a heterogeneous
rectangular unit cell $\Omega_{i}=[0,L_{x}]\times[0,L_{y}]$, our
goal is to predict its mechanical response under given microstructure
and boundary conditions (BCs). As introduced before, we use LSM to
convert unit-cell microstructure into a continuous distance field
$\phi(x,y)$. Similarly, to convert a given BCs into a continuous
field, we first divide $\Omega_{i}$ into a coarse mesh of $N_{c}\times N_{c}$
nodes. Then, we ignore the internal microstructure and treat the entire
unit cell as linear and homogeneous. The elastic modulus of such imaginary
cell is set to the volumetric average of SiC and Al. Next, we use
FEM to calculate the responsive displacement field $u_{coarse}$,
and numerically calculate the coarse strain field $\epsilon_{coarse}$.
As can be seen, the BCs are encoded into the coarse displacement and
strain fields. The FNO input tensor $t(x,y)$ can be prepared as follows,
\[
t(x,y)=[u_{coarse},\epsilon_{coarse},\phi(x,y),x,y]
\]
The direct output of FNO is set to residual field $u_{res}$, which
is the discrepancy between $u_{coarse}$ and the true response of
the heterogeneous composite. As a result, the true displacement solution
$u_{fine}$ can be obtained via $u_{fine}=u_{coarse}+u_{res}$. The
fine strain field $\epsilon_{fine}$ is calculated from $u_{fine}$,
and they form the final output of the FNO model. For the convenience
of subsequent description, we use $\mathcal{H}$ to denote the unit-cell
FNO model, namely, $(u_{fine},\epsilon_{fine})=\mathcal{H}(g_{i};\phi)$.

The above construction strategy of building-block neural network for
unit cell has the following advantages. First, the employment of residual
learning can significantly simplify the training of FNO. Since the
target of FNO is set to the discrepancy between true solution and
coarse field $u_{coarse}$, we have excluded a large portion of non-periodic
modes, leading to reduced bias and faster loss decay. Second, the
usage of LSM eliminates the spatial discontinuity, leading to highly
concentrated low-frequency modes. Since the discontinuities at matrix-reinforcement
interface is prone to generate large amounts of high-frequency mode,
thereby exciting Gibbs oscillations. LSM can effectively smoothen
the interface and get rids of these undesired high frequencies.

\subsection{Construction of Neural-Swarm for large-scale problem}

For large-scale composite domain with rich microscopic geometric features
and strong material nonlinearity, conventional modeling method such
as FEM usually requires prohibitively high computational cost for
high physical fidelity. Here, we use a domain decomposition scheme
to partition the large-scale composite into subdomains of typical
microstructure features. Since we have constructed building-block
FNO models to describe the behaviors of typical microstructure, we
can integrate these building blocks into an FNO swarm according to
the spatial adjacency of the subdomains. Under arbitrarily enforced
Dirichlet BCs, the mechanical response of the large-scale composite
structure will be determined by collective inference on the FNO swarm
using schwarz alternating method. Moreover, we will give a rigorous
analysis about the convergence of FNO swarm. Based on the result of
microstructural analysis, we partition a large composite domain $\Omega$
into $N$ subdomains, i.e., $\Omega=\cup_{i=1}^{N}\Omega_{i}$, so
that each subdomain $\Omega_{i}$ can be represented by a unique FNO
model. To make the building-block FNOs exchange information, the adjacent
subdomains should overlap a little so that they share common sampling
points. As a result, the FNOs of subdomains can be assembled by neuron
sharing, which essentially becomes a swarm of FNOs, as illustrated
in Fig.6. Compared to single FNO model, FNO swarm enables flexible
integration of arbitrary number of FNO models, which is of great advantage
for the simulation of irregular domain. The goal of collective inference
is to achieve a systematic equilibrium among all FNO models so that
the global FNO swarm converges to a stable state.

\begin{figure}[H]
\centering \includegraphics[width=0.95\textwidth]{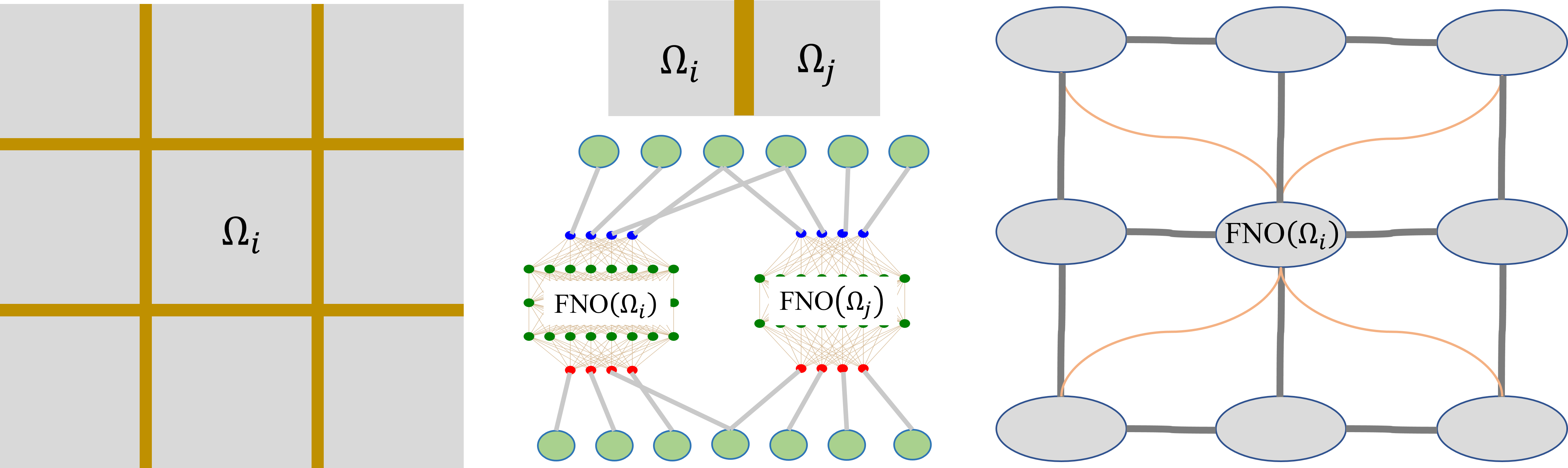}
\caption{(a)Schematic illustration of domain decomposition with overlapping.
(b)Neuron sharing among FNOs of adjacent subdomains. (c)the FNO swarm
formed by integrating the FNOs corresponding to the subdomains.}
\label{fig:domain_decomp} 
\end{figure}

The collective inference of the FNO swarm is realized by Schwarz alternating
iterations. In the traditional overlapping schwarz alternating method,
iteration usually starts with a zero displacement field. For highly
nonlinear problems, it usually leads to extremely slow dissipation
of low-frequency, and is very prone to local optimal. To overcome
this issue, we start with an initial guess of $\mathbf{u}_{global}^{(0)}$
directly obtained from FEM. Specifically, we create a coarse mesh
$\Gamma_{H}$ on the simulation domain $\Omega$ with $N_{init}\times N_{init}$
nodes. For this initialization FEM calculation, we ignore the internal
microstructure and regard $\Omega$ as an effectively isotropic elastic
medium whose modulus is the volumetric average of the constituent
phases. Denoting the FEM solution by $\mathbf{U}_{H}$, we employ
a bicubic interpolation operation $\Upsilon:\Gamma_{h}\rightarrow\Gamma_{m}$
to obtain the initial displacement field $\mathbf{u}_{global}^{(0)}$,
i.e., 
\[
\mathbf{u}_{global}^{(0)}(\mathbf{x})=\Upsilon(\mathbf{U}_{H}),\quad\forall\mathbf{x}\in\Omega
\]
For $k$-th iteration, the response field from previous iteration
$u_{global}^{(k-1)}$ is dispatched to each subdomain and for each
subdomain, we only consider the responsive values on the exterior
boundary via 
\[
g_{i}^{(0)}=\mathbf{u}_{global}^{(0)}|_{\partial\Omega_{i}}
\]
These exterior values are used as Dirichlet BCs for the FNO of that
subdomain, which is then activated to to predict the responsive field
within. Because each subdomain $\Omega_{i}$ is handled independently,
the solutions of adjacent subdomains will lead to inconsistency within
the overlapping area before convergence is reached. To generate a
smooth update for next iteration, we introduce the partition of unity
method to fuse the solutions from adjacent subdomains. Specifically,
we define a 2D Hamming window as a spatial non-negative weight function
$W_{i}(x,y)$ within the local coordinate system of each unit cell
$\Omega_{i}$. For a point $(x,y)\in[0,L_{x}]\times[0,L_{y}]$ in
$\Omega_{i}$, $W_{i}(x,y)$ is defined as 
\[
W_{i}(x,y)=[\alpha-\beta cos(\frac{2\pi x}{L_{x}})]\times[\alpha-\beta cos(\frac{2\pi y}{L_{y}})]
\]

where $\alpha=0.54$, $\beta=0.46$. This function reaches a maximum
of 1 at the geometric center of the unit cell and smoothly decays
in a second-order manner to near 0 toward the unit cell boundaries.
A smooth response field for next iteration, $u_{global}^{(k)}$, can
be expressed as the weighted average of all subdomain solutions, 
\[
u_{global}^{(k)}=\frac{\sum_{i=1}^{N}W_{i}\cdot u_{fine,i}^{(k)}}{\sum_{i=1}^{N}W_{i}}
\]

\begin{figure}[H]
\centering \includegraphics[width=0.95\textwidth]{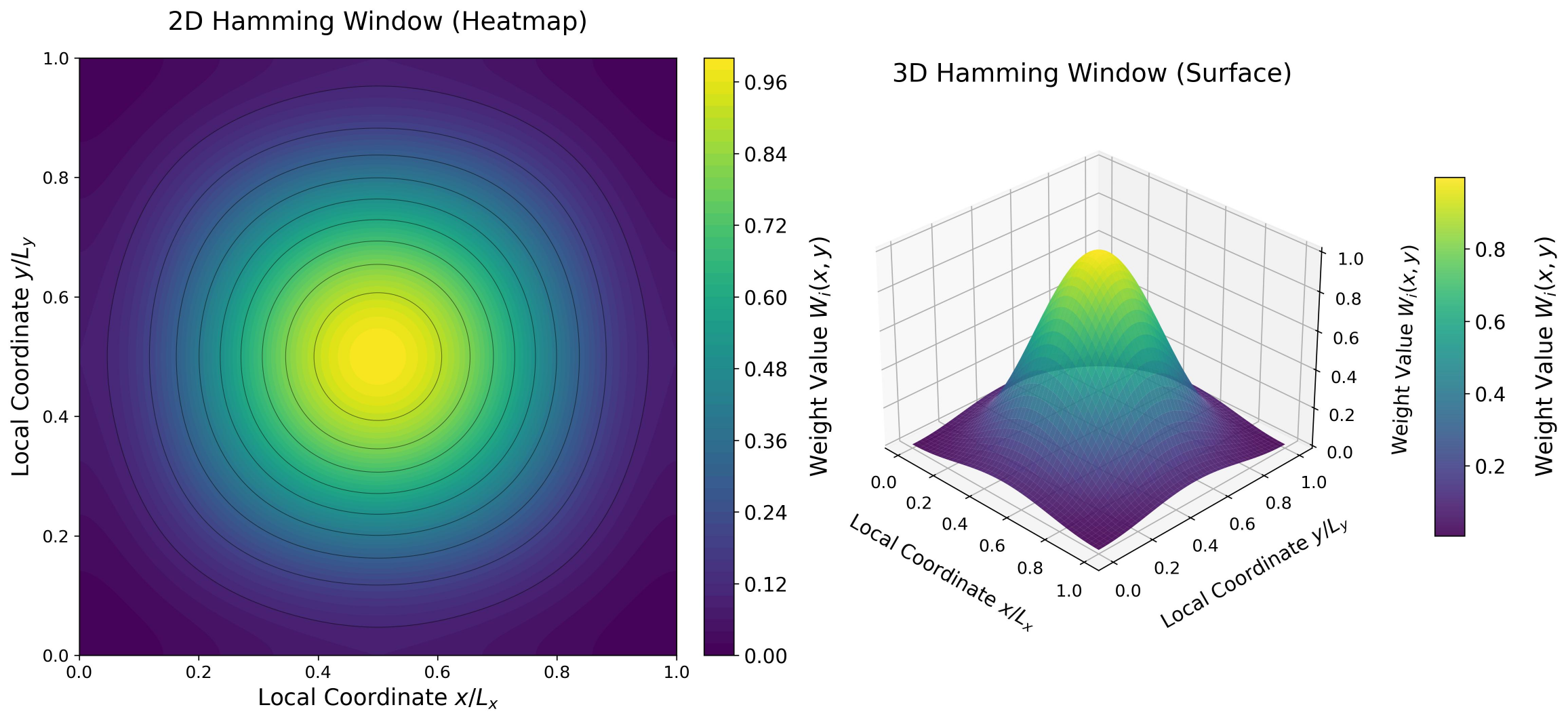}
\caption{Distribution diagram of the 2D Hamming window weighting function}
\label{fig:hamming} 
\end{figure}

The relative error of the global displacement field between two adjacent
iterations is calculated to determine whether convergence has been
reached. We use the $L^{2}$ norm to define the convergence tolerance,
\[
\frac{||u_{global}^{(k)}-u_{global}^{(k-1)}||_{L^{2}(\Omega)}}{||u_{global}^{(k-1)}||_{L^{2}(\Omega)}}<\epsilon
\]

After sufficient iterations, the procedure terminates when the relative
error is less than a predefined threshold. Under this circumstance,
the temperature fields of adjacent subdomains become consistent within
the overlapping regions and the subdomain FNOs are systematically
synchronized. The resulted global response field is the final answer
to the original problem. To analyze the error evolution during the
Schwarz iterations, let $u$ be the true solution over the global
domain $\Omega$. In the $k$-th iteration step, the global displacement
field is denoted by $u_{global}^{(k)}$, and the global absolute error
at this stage can be defined as 
\[
E^{(k)}=u_{global}^{(k)}-u
\]
The global absolute error consists of two sources, the subdomain inference
error and the propagated iterative error from the overlapping domains.
For subdomain $\Omega_{i}$, given boundary conditions $g_{i}$, let
the governing partial differential operator be $\mathcal{F}(g_{i})$,
and the solution predicted by its FNO be $u_{fine}=\mathcal{H}(g_{i};\phi)$,
where $\mathcal{H}$ represents the mapping relation of the FNO as
mentioned before. We assume that there is an upper bound $\delta$
such that for any admissible boundary condition $g_{i}$, there is:
\[
||\mathcal{H}(g_{i};\phi)-\mathcal{F}(g_{i})||\leq\delta
\]
In the $k$-th iteration update, we introduce the normalized partition
of unity weight function 
\[
\bar{W}_{i}(x)=\frac{W_{i}(x)}{\sum_{j=1}^{N}W_{j}(x)}
\]
The global solution is constructed by the weighted average of all
subdomain solutions. To isolate the error sources, we define an ideal
subdomain update field $u_{exact,i}^{(k)}$, which is the exact solution
of the governing PDE under the boundary conditions $g_{i}^{(k-1)}$
of subdomain $\Omega_{i}$ extracted of the global solution from $(k-1)$th
iteration, 
\[
u_{exact,i}^{(k)}=\mathcal{F}(g_{i}^{(k-1)})
\]
As a result, the exact solution of the governing PDE and the update
from Schwarz iteration can be written as 
\[
u_{exact}^{(k)}=\sum_{i=1}^{N}\bar{W_{i}}u_{exact,i}^{(k)},\ u_{fine,i}^{(k)}=\mathcal{H}(g_{i}^{(k-1)};\phi)
\]

Now consider the global absolute error $E^{(k)}=u_{global}^{(k)}-u$:
\[
E^{(k)}=u_{global}^{(k)}-u_{exact}^{(k)}+u_{exact}^{(k)}-u
\]

Taking the $H^{(1)}(\Omega)$ norm on both sides of the equation,
and applying the triangle inequality: 
\[
||E^{(k)}||_{H^{1}(\Omega)}\leq||u_{global}^{(k)}-u_{exact}^{(k)}||_{H^{1}(\Omega)}+||u_{exact}^{(k)}-u||_{H^{1}(\Omega)}
\]

For the right-hand side of the above inequality, $||u_{exact}^{(k)}-u||_{H^{1}(\Omega)}$
term represent the error from Schwarz iteration. According to the
convergence theory of overlapping Schwarz method while in solving
for elliptic PDEs, the Schwarz iteration is a contraction mapping,
i.e., 
\[
||u_{exact}^{(k)}-u||_{H^{1}(\Omega)}\leq\rho||u_{exact}^{(k-1)}-u||_{H^{1}(\Omega)}=||E^{(k-1)}||_{H^{1}(\Omega)}
\]

where $\rho\in(0,1)$ is a constant depending on the width of overlapping
domain. $|u_{global}^{(k)}-u_{exact}^{(k)}||_{H^{1}(\Omega)}$ is
the perturbation brought by the building-block FNOs to the global
field. Now let us turn our attention back to the subdomain and investigate
how the error is originated and how it affects the global solution.
Denoting the error within the subdomain by $e_{i}^{(k)}=u_{fine,i}^{(k)}-u_{exact,i}^{(k)}$,
it is well bounded by 
\[
||e_{i}^{(k)}||_{H^{1}(\Omega)}=||u_{fine,i}^{(k)}-u_{exact,i}^{(k)}||_{H^{1}(\Omega)}\leq\delta
\]
So the global perturbation term can be reformulated as 
\[
||u_{global}^{(k)}-u_{exact}^{(k)}||_{H^{1}(\Omega)}=||\sum_{i=1}^{N}\bar{W_{i}}e_{i}^{(k)}||_{H^{1}(\Omega)}
\]
According to the definition of $H^{1}$ norm, 
\[
||\bar{W_{i}}e_{i}||_{H^{1}(\Omega)}=||\bar{W_{i}}e_{i}||_{L^{2}}^{2}+||e_{i}\nabla\bar{W_{i}}+\bar{W_{i}\nabla e_{i}}||_{L^{2}}^{2}
\]
Again, with the application of triangle inequalities, we can obtain
\[
||\bar{W_{i}}e_{i}||_{H^{1}(\Omega)}\leq||\bar{W_{i}}||_{L^{\infty}}^{2}||e_{i}||_{L^{2}}^{2}+2||\bar{W_{i}}||_{L^{\infty}}^{2}||\nabla e_{i}||_{L^{2}}^{2}+2||\bar{\nabla W_{i}}||_{L^{\infty}}^{2}||e_{i}||_{L^{2}}^{2}
\]
Since the amplitude of the normalized weight function $||\bar{W_{i}}||_{L^{\infty}}\leq1$,
and $|e_{i}||_{H^{1}(\Omega)}^{2}=||e_{i}||_{L^{2}}^{2}+||\nabla e_{i}||_{L^{2}}^{2}\leq\delta^{2}$,
rearranging the above equation leads to 
\[
||\bar{W_{i}}e_{i}||_{H^{1}(\Omega)}\leq C_{w}\delta
\]
where $C_{w}$ is a constant. Based on this error bound of subdomain,
we can now determine the bound of the global perturbation, $||\sum_{i=1}^{N}\bar{W_{i}}e_{i}^{(k)}||_{H^{1}(\Omega)}$.
According to the domain decomposition scheme discussed above, any
spatial point is affected by $N_{c}=4$ adjacent subdomains at most.
Therefore, when computing the norm of global perturbation, the number
of cross terms is strictly bounded by $N_{c}$. Utilizing the finite
sum lemma for spatial domain decomposition, we have 
\[
||\sum_{i=1}^{N}\bar{W_{i}}e_{i}^{(k)}||_{H^{1}(\Omega)}^{2}\leq N_{c}\sum_{i=1}^{N}||\bar{W_{i}}e_{i}^{(k)}||_{H^{1}(\Omega)}
\]
\[
||\sum_{i=1}^{N}\bar{W_{i}}e_{i}^{(k)}||_{H^{1}(\Omega)}\leq\sqrt{N_{c}}max_{i}||\bar{W_{i}}e_{i}||_{H^{1}(\Omega_{i})}\leq\sqrt{N_{c}}C_{w}\delta
\]

Going back to $||E^{(k)}||_{H^{1}(\Omega)}$, we can immediately obtain
the error propagation relation, 
\[
||E^{(k)}||_{H^{1}(\Omega)}\leq\rho||E^{(k-1)}||_{H^{1}(\Omega)}||+C\delta
\]
As $k\rightarrow\infty$, we can obtain the following error limit
\[
lim_{k\rightarrow\infty}||E^{(k)}||\leq\frac{C\delta}{1-\rho}
\]

According to above analysis, we can see that, for the error introduced
by the inference of building-block FNOs of the subdomains, the Schwarz
iteration controls the error propagation, preventing its unbounded
accumulation and divergence. Moreover, the final error of the global
solution is approximately proportional to the prediction error $\delta$
of the local FNO model. As along as the prediction error of the unit-cell
FNO is sufficiently small, the error of the global solution from Schwarz
method can be controlled within a certain threshold. Finally, it also
manifests the significance of initialization before Schwarz iteration.
The application of coarse-mesh FEM initialization provides an excellent
$||E^{(0)}||$, which can strongly accelerate convergence.

\section{Numerical Experiments}

\subsection{Building-block FNOs for SiC-Al unit cells}

For SiC particle-reinforced aluminum matrix composites, the size and
spatial arrangement of the SiC particles are the primary factors influencing
the overall mechanical behavior of the composite. Here, we consider
two typical microstructural configurations: a unit cell containing
a single large SiC particle (referred to as unit cell A) and a unit
cell containing multiple smaller SiC particles (unit cell B), as shown
in Fig.8(a)(b). These two cells represent the scenarios of fine and
coarse dispersion of reinforcement phase, respectively. Both unit
cells are squares with dimensions of $100\times100$, where Unit Cell
A contains a centrally located SiC particle with a radius of 30, whereas
Unit Cell B contains three uniformly distributed particles with a
radius of 10.

The FNO for each unit cell is configured with 12 modes, a hidden layer
width of 64, and 4 Fourier layers. Its input features consist of an
8-channel tensor $\mathbf{a(x)}$, which includes the initial displacement
field $\mathbf{u}_{coarse}=[u_{0}(\mathbf{x}),v_{0}(\mathbf{x})]$,
the initial strain field $\mathbf{\epsilon}_{coarse}=[\epsilon_{xx,0},\epsilon_{yy,0},\gamma_{xy,0}]$
specifically tailored for the nonlinearity of the composite system,
the spatial coordinates $\mathbf{x}=[x,y]$, and the signed distance
field (SDF) $\phi(\mathbf{x})$. The output of each FNO is a 5-channel
tensor $\mathbf{b}$, encompassing the displacement field $u_{fine}$
and the strain field $\mathbf{\epsilon}_{fine}$ on the fine mesh.
As previously mentioned, $u_{coarse}$ and $\mathbf{\epsilon}$ are
interpolated from the coarse-mesh linear FEM simulation results of
the hypothetical homogeneous equivalent, and $\phi(\mathbf{x})$ is
obtained via the level set method (LSM) introduced earlier. For a
unit cell with multiple reinforcing particles, its SDF at point $\mathbf{x}=(x,y)$
is determined as the difference between the distance from $\mathbf{x}$
to the nearest particle center and the radius of that particle, namely:
\[
\phi(\mathbf{x})=R-min||\mathbf{x-x_{c}}||_{2}
\]
where $R$ is the radius of the particle closest to the current position,
and $\mathbf{x}_{c}$ is the center of that particle. Fig.8(c)(d)
illustrates the calculated SDF of the unit cells. It can be observed
that the originally discontinuous microstructure is transformed into
a continuous field, which will significantly reduce the training difficulty
of the FNO without introducing any additional information. 
\begin{figure}[htbp]
\centering \includegraphics[width=0.85\textwidth]{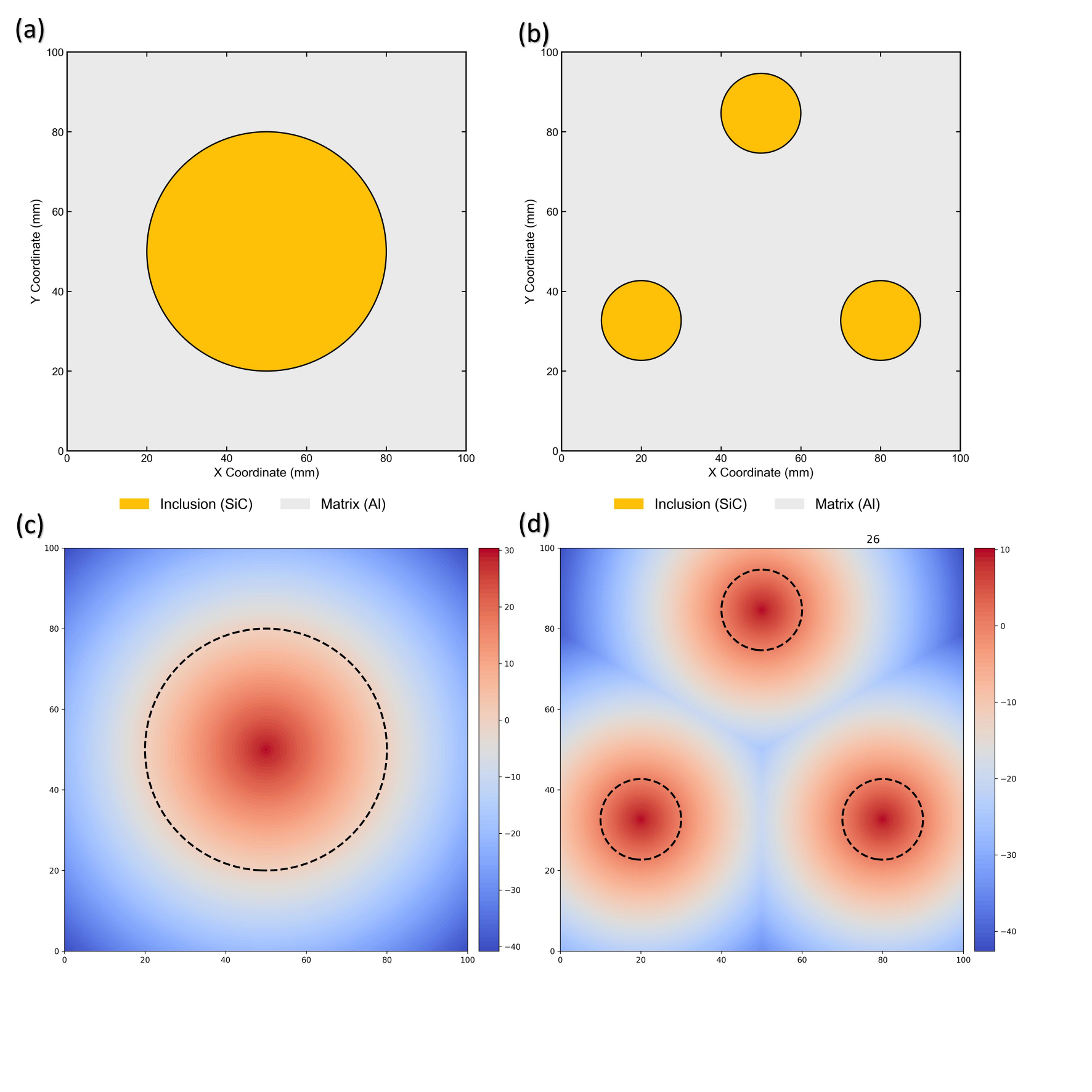}

\caption{Microstructural configurations of unit cell A and unit cell B, along
with their corresponding signed distance fields (SDFs) calculated
via the level set method (LSM).}
\label{fig:UnitCellConfigSDF} 
\end{figure}

As briefly discussed before, to collect the mechanical behaviors for
training the unit-cell FNOs, we discretize each unit cell domain into
$50\times50$ nodes and generate fine mesh of triangular elements.
We repetitively employ FEM on the fine mesh by randomly varying the
boundary conditions (BCs) and gather 30000 data samples as the training
set for each of them. During training, we employ the $H_{1}$ loss
function, which includes the mean squared error of the predicted displacements
and strains, as well as the spatial gradient loss of the displacement
field, given by: $Loss=MSE(u,v,\epsilon_{xx},\epsilon_{yy},\epsilon_{xy})+\beta MSE_{grad}(u,v)$,
where $\beta=0.1$. Additionally, we use the AdamW optimizer with
an initial learning rate of 0.001, dynamically reducing it via a cosine
annealing strategy. At the end of training, the value of $Loss$ can
converge to below $10^{-5}$.

To validate the prediction accuracy of trained unit-cell FNOs intuitively,
we have designed the following test cases, whose Dirichlet BCs are
sufficiently complex and different from the training set. Given that
the unit cell is a square domain, we define a normalized coordinate
$s\in[0,1]$ along each edge, and apply non-uniform displacement boundaries
$\mathbf{u}=\{u,v\}$ to the edges. To ensure the Dirichlet BCs on
different edges are consistent at the corner points, we enforce a
zero displacement on each corner. Specifically, on the left edge,
$\ensuremath{u=-A\cdot\Phi_{1}(s)}\text{, }\ensuremath{v=0}$, on
the right edge, $\ensuremath{u=B\cdot\Phi_{2}(s)}\text{, }\ensuremath{v=A\cdot\Phi_{1}(s)}$,
on the top edge, $\ensuremath{u=B\cdot\Phi_{3}(s)}\text{, }\ensuremath{v=B\cdot\Phi_{4}(s)}$,
on the bottom edge, $\ensuremath{u=-A\cdot\Phi_{1}(s)}\text{, }\ensuremath{v=0}$.
Fig.9 plots the patterns of $\Phi_{1}(s)\sim\Phi_{4}(s)$ along cell
edges.

\begin{figure}[H]
\centering \includegraphics[width=0.95\textwidth]{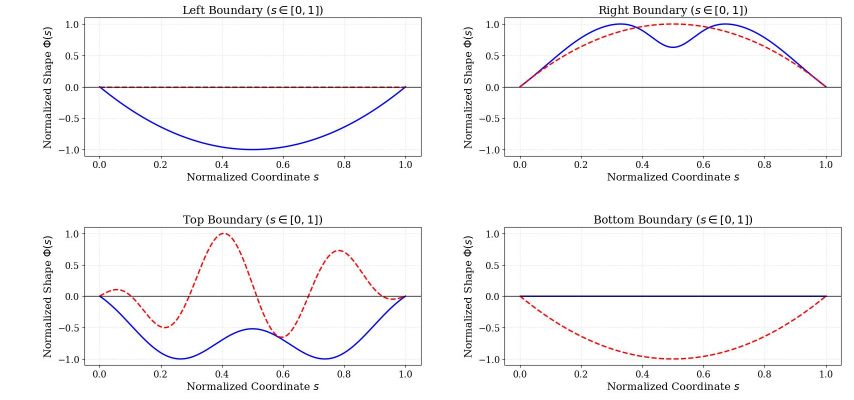}
\caption{Plots of $u$ and $v$ of Dirichlet BCs along each edge, with $A=1$
and $B=1$.}
\end{figure}

Here, $A$ and $B$ are random parameters controlling the magnitude
of the applied Dirichlet BCs. When $A$ and $B$ are sufficiently
small, the overall magnitude of the Dirichlet boundary conditions
is small, the SiC-Al composite is expected to exhibit linear behavior.
When $A$ and $B$ are sufficiently large, the strain state of the
unit cell will might exceed its elastic limit, thereby leading to
nonlinear behavior. To investigate the performance of the building-block
FNOs in both scenarios, we prepared two cases. Case 1 is performed
on unit cell B which focuses on the linear scenario, where $A=0.02$
and $B=0.1$; Case 2 is performed on unit cell A which focuses on
the nonlinear scenario, where $A=B=2.5$. We first employ classical
FEM to obtain the displacement and strain fields for both cases, and
calculate the displacement magnitude and equivalent strain at each
node as ground truths.

Fig.10 displays the FNO-predicted and FEM-calculated displacement
fields and equivalent strain fields for Unit Cell B under linear elastic
conditions. It can be seen that the predicted displacement and strain
fields are of similar patterns as that of the ground truths. Quantitatively,
the error maps prove that both the average and maximum prediction
errors of the displacement and strain results are rather small, suggesting
that trained FNO of unit cell B behaviors well for this test case.
Closer inspection tells us that the maximum error occurs at the exterior
boundary where the Dirichlet BCs changes drastically. This is understandable,
since the abruptly changing BC typically leads to strong deformation
gradient and complex strain state. This features can only be properly
dealt with by including sufficient amounts of high-frequency mode.
However, the high-frequency truncation effect is inherent for FNO
modeling, resulting in over-smoothing at locations with drastic field
transitions.

\begin{figure}[H]
\centering \includegraphics[width=0.95\textwidth]{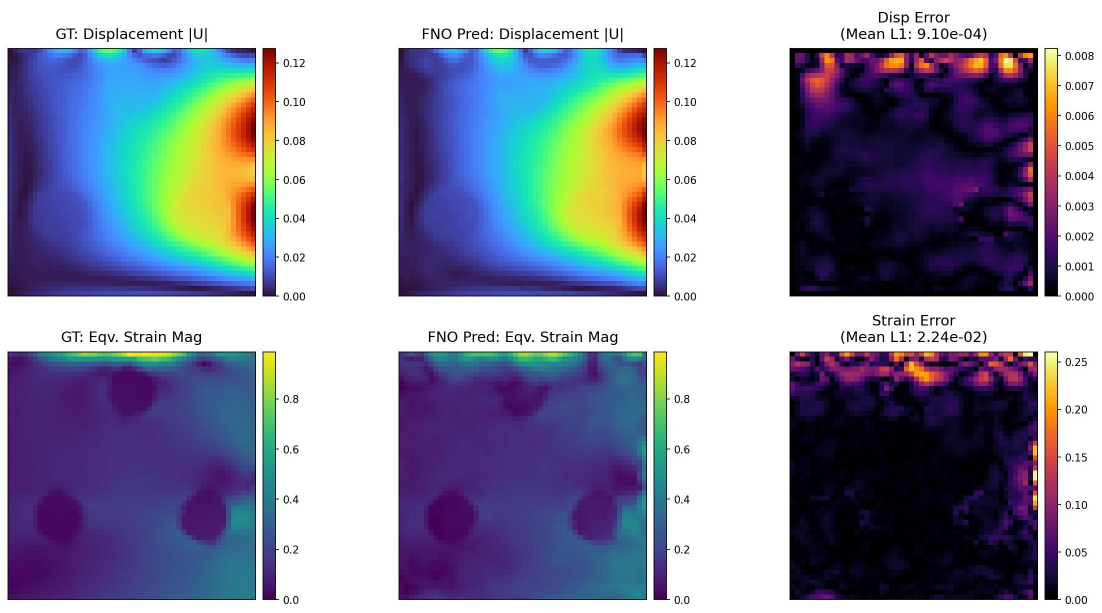}
\caption{Test case for the trained FNO model of unit cell B under the Dirichlet
BCs of $A=0.02,B=0.1$.The first column shows the displacement magnitude
and Von‑Mises strain obtained from FEM calculations, the second column
shows those predicted by FNO, and the third column shows the discrepancies
between these two. Here, displacement magnitude is calculated via
displacement $|U|=\sqrt{u^{2}+v^{2}}$, equivalent strain is obtained
via $\epsilon_{vm}=\sqrt{\varepsilon_{xx}^{2}+\varepsilon_{yy}^{2}-\varepsilon_{xx}\varepsilon_{yy}+3\varepsilon_{xy}^{2}}$.}
\end{figure}

Fig.11 illustrates the FNO predictions for unit cell A under nonlinear
conditions ($A=B=2.5$) and the corresponding ground truths from FEM
calculation. As can be seen, the FNO model correctly reproduces the
overall pattern of the mechanical responses under such BCs, suggesting
that the material nonlinearity of SiC-Al composite has been reasonably
captured by FNO. Again, the largest error source still comes from
the exterior boundary where large displacement variation is imposed.
However, compared to the linear case of small-magnitude Dirichlet
BCs, an appreciable increment in both average and maximum errors can
be observed. It indicates that once nonlinear mechanical behavior
is introduced, the FNO's prediction quality noticeably degrades, confirming
that the entanglement of material nonlinearity and complex boundary
conditions significantly increases the difficulty of neural network
training as well as generalization. Fortunately, as we will see in
the following numerical experiments, the current level of prediction
accuracy of the unit-cell FNOs is sufficient be employed as building
blocks for large-scale composite simulations. This is mainly attributed
to the correctness of the predicted field patterns and the average
magnitudes.

\begin{figure}[H]
\centering \includegraphics[width=0.95\textwidth]{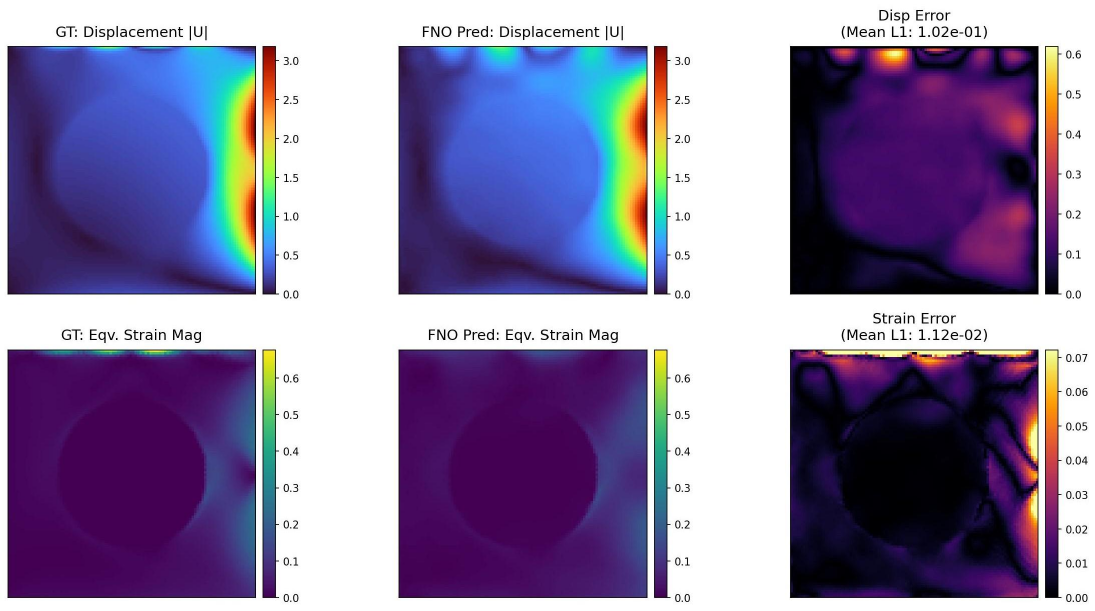}
\caption{Test case for the trained FNO model of unit cell A under large deformation
that leads to nonlinear material behavior. Again, the first column
is the FEM results of displacement magnitude field and equivalent
strain field, the second column is the FNO prediction results, while
the third column is the discrepancy maps.}
\label{fig:nl_disp_a} 
\end{figure}

Finally, to demonstrate the statistical robustness of the trained
unit-cell FNOs, we randomly generate 100 sets of Dirichlet BCs which
are not seen during the training process, and utilize the trained
FNO model of unit cell A to predict the mechanical responses under
these BCs. Meanwhile, FEM is employed to calculate the corresponding
ground truth for each BC case. Then, we calculate the point-by-point
discrepancy between the FNO prediction and the ground truth and measure
the average error across the entire unit cell domain. Fig.12 shows
the average errors of two measures, i.e., the displacement magnitude
and the equivalent strain, with respect to the maximum values of the
imposed Dirichlet BCs. As can be seen, the average prediction error
of the trained FNO model is grossly linearly dependent on the magnitude
of the applied Dirichlet BCs, suggesting an approximately constant
relative prediction error. Moreover, the prediction error of equivalent
strain is generally larger than the error of displacement magnitude,
which is understandable since the equivalent strain is derived from
the predicted displacement field.

\begin{figure}[H]
\centering \includegraphics[width=0.95\textwidth]{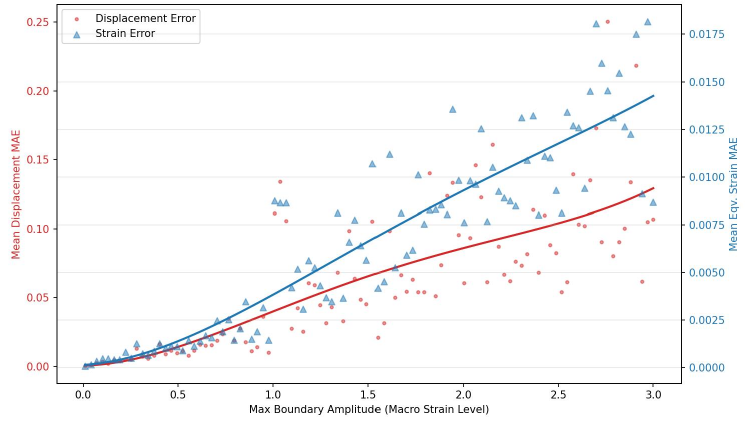}
\caption{Variation of the average errors of displacement and strain fields
predicted by the trained FNO with respect to the maximum boundary
displacement. The red data points represent the mean squared error
of displacement magnitude, the blue data points represent the mean
squared error of the equivalent Von‑Mises strain, and the horizontal
axis is the maximum value of the applied Dirichlet BCs.}
\end{figure}

\subsection{Frequency-domain analysis}

Since the microstructure of composite material has discontinuous interface
between the reinforcement and matrix phases, which usually requires
infinite amount of high-frequency modes in the spectral domain to
resole the discontinuity. To avoid such difficulty, we have employed
the LSM to represent the original discontinuous microstructure by
a smooth SDF field. The benefit of this operation can be appreciated
from Fig.13(a)-(c), where power spectral density (PSD) map is obtained
for both representations. Compared to the original discontinuous microstructure,
the spectral power of the SDF field is highly concentrated at the
low-frequency end, and the high-frequency modes decay very fast. This
means that we can retain the major physical information if the high-frequency
modes are truncated. To confirm this assertion, we can perform Fast
Fourier Transformation (FFT) on both the SDF field and the original
discontinuous microstructure and set the high-frequency coefficients
to zero to just keep the first 16 modes. Then, we transform the truncated
frequency signals back to the physical space via inverse FFT to reconstruct
the input data. To assess the quality of the reconstructions, we horizontally
section the original and the reconstructed fields through the centers.
As can be seen from Fig.13(d)-(f), the low-frequency reconstruction
of the SDF field matches perfectly with the its original input. In
contrast, for the reconstruction directly from the original discontinuous
microstructure, severe Gibbs oscillations occurs at the interfacial
vicinity, as depicted in Fig.13(g)-(i). This comparison justifies
the usage of the smooth SDF as a continuous representation of the
composite microstructure for better alignment with FNO modeling.

\begin{figure}[H]
\centering \includegraphics[width=0.95\textwidth]{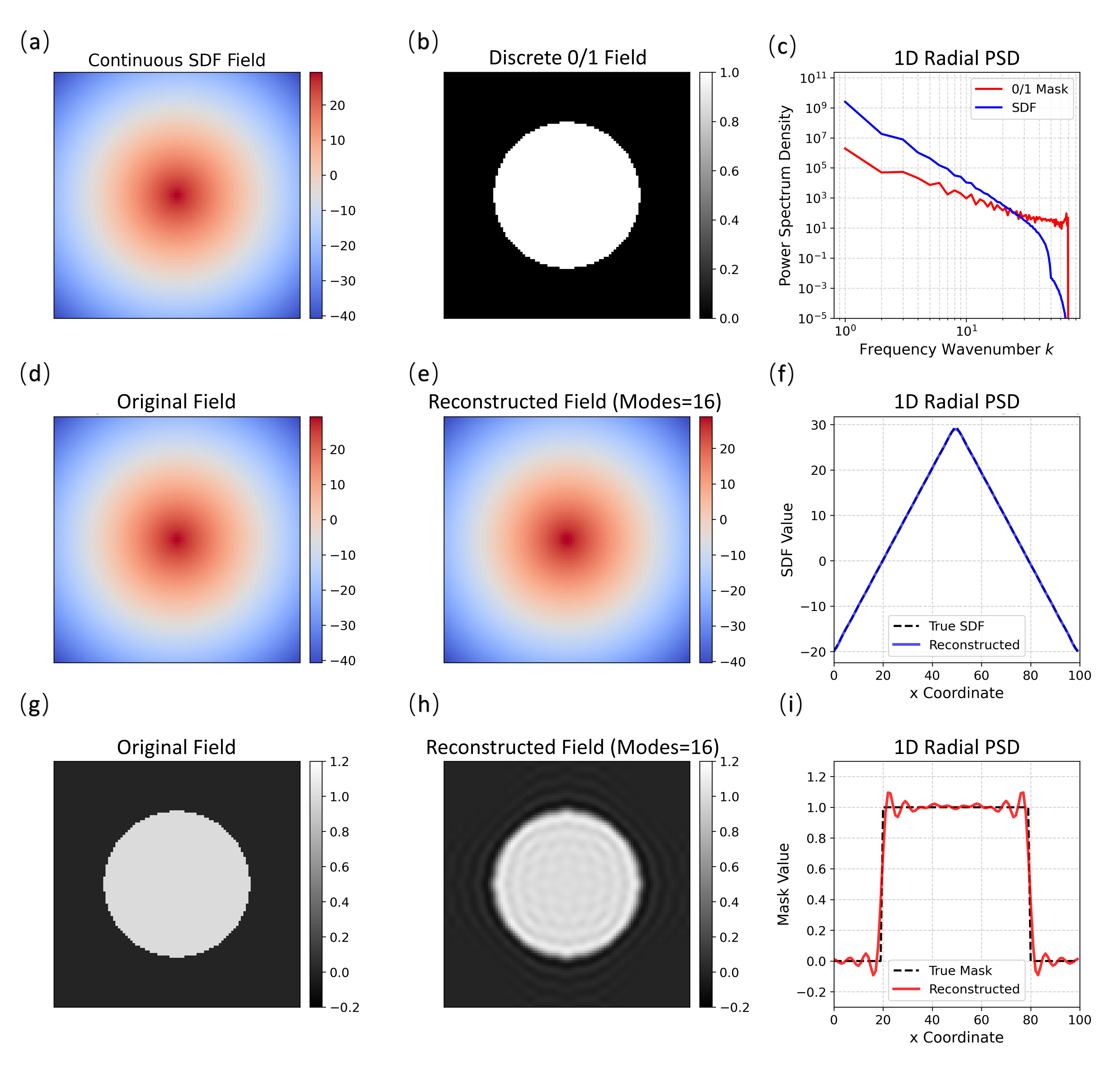}
\caption{Spectral analysis of the discontinuous microstructure and the converted
SDF. (a) The continuous SDF field of unit cell A. (b) the binarized
field of the original discontinuous microstructure.(c)the power spectra
of (a) and (b) through FFT. (d)-(e)the SDF and the reconstruction
of SDF via the lowest 16 frequencies. (f) the field values along the
horizontal line passing through the centers of the original and reconstructed
SDFs. (g)-(h)the binarized discontinuous microstructure and its construction
via the lowest 16 frequencies. (i) the field values along the horizontal
line passing through the centers of the binarized and reconstructed
microstructures.}
\end{figure}

As we have discussed previously, the goal of the unit-cell FNO is
to learn the residual between the ground truth and the coarse result,
namely, $u_{res}=u_{true}-u_{coarse}$. Fig.14 shows typical spatial
distributions of the displacement and strain residuals after the coarse-mesh
initialization, as well as their PSDs. It is evident that the residual
is largely confined to the vicinity of the reinforcement-matrix interface.
Further, from the power spectral density, we can see that, in the
low-frequency range, the coarse-mesh initialization highly overlaps
with the ground truth, indicating that it has grossly captured the
overall pattern of deformation field. When entering the high-frequency
range, the power density of the coarse-mesh initialization decays
sharply, causing it to completely deviate from the ground truth. This
indicates that the coarse initialization doesn't contain the high-frequency
information, which is left for the FNO model to learn. The PSD plot
also shows that the strain field derived from the coarse-mesh initialization
fails to capture the spectral information even from the low-frequency
end, which indicates that the FNO model has to learn almost the entirety
spectrum of the strain field. This explains why the prediction error
of the unit-cell strain field is much larger than that of the displacement
field.

\begin{figure}[H]
\centering \includegraphics[width=0.95\textwidth]{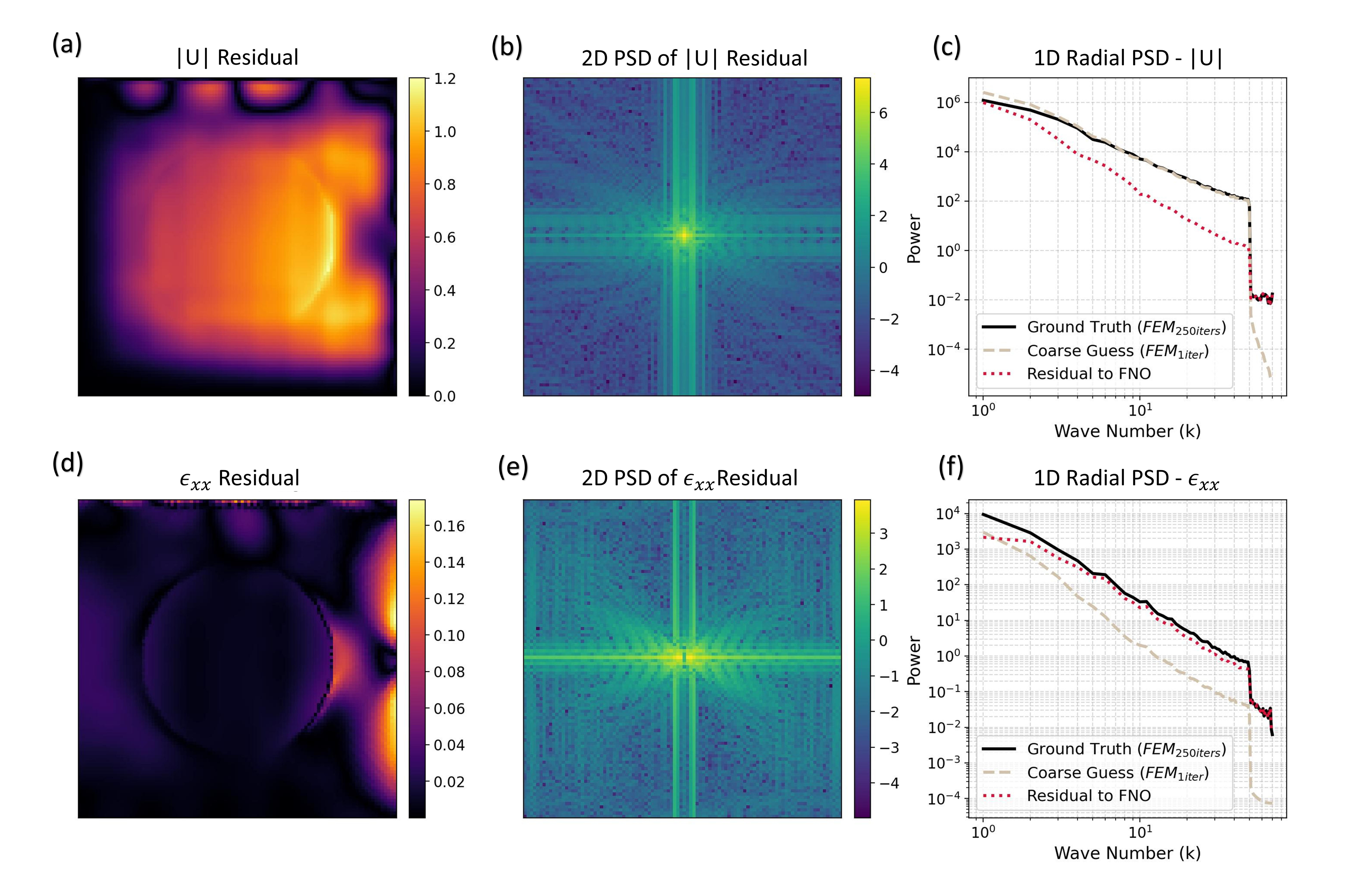}
\caption{Spectral analysis of the residual fields.(a)residual field of displacement
magnitude $|U|$. (b)2D power spectrum of (a). (c)1D radial power
spectrum of (a). (d)residual field of strain field $\epsilon_{xx}$.
(e) 2D power spectrum of (d). (f) 1D radial power spectrum of (d).
The 2D power spectra are obtained via the 2D FFT of the residual fields.
The 1D radial PSD is obtained by performing azimuthal averaging of
the two-dimensional power spectral density over concentric circular
rings centered at the zero-frequency point. In (c) and (f), the coarse
guess is obtained by linear FEM over a 10\texttimes 10 mesh, which
is then upsampled to a 100\texttimes 100 resolution using bicubic
interpolation, while the ground truth is obtained by nonlinear FEM
after 250 iterations directly over a 100\texttimes 100 mesh.}
\end{figure}

\subsection{FNO swarm for large-scale Al-SiC composite}

As discussed before, the FNO swarm is constructed by integrating the
above unit-cell FNOs according to the microstructural configuration
of the simulation domain to be solved. It is intrinsically capable
of dealing with composite domains of arbitrary length scales. In the
following, we will thoroughly test the computational efficiency and
physical fidelity of the FNO-swarm method solving for various large-scale
simulation domains of Al-SiC composite under various geometric shapes
and boundary conditions.

\subsubsection{Composites of single microstructure feature}

To start with, let us consider two simulation cases for relatively
simple simulation domains of pure unit cell A and pure unit cell B,
respectively. For each case, the entire simulation domain contains
$50\times50$ unit cells. The boundary condition follows the same
logic as discussed before, and $A$ and $B$ are set to 4.3. To determine
the mechanical response of the simulation domain by the FNO-swarm
method, we first integrate the building-block FNOs according to the
adjacency of the unit cells within the domain. Then, we initialize
the input neurons of the building-block FNOs as follows. We assume
the simulation domain is purely Al matrix and construct an extremely
coarse mesh over it. As a simple approach, the corner points of the
unit cells can be used as the nodal point and the coarse mesh simply
consists of $50\times50$ quadrilateral elements, which can be calculated
by FEM with negligible time cost. The calculated displacement is then
assigned to the unit-cell corner points and bicubic interpolation
is utilized to determine the displacement values on the edges of each
unit cell.

Knowing the displacement field values on the exterior edges, we can
invoke the Schwarz alternating iteration to iteratively call the building-block
FNO and update the displacement field of the simulation domain, so
that the predicted displacement fields of adjacent unit cells are
consistent within the shared regions. For comparison, we also employ
FEM to directly calculate the displacement field for each simulation
domain under the same Dirichlet BCs. For FEM calculation, the entire
simulation domain is directly meshed into fine triangular elements.
Fig. 15 demonstrates the microstructural configuration of the two
cases, as well as the calculated displacement fields. As can be seen,
compared to the ground truths, the overall patterns of the displacement
fields have been accurately reproduced by the FNO swarm. Similar as
the prediction error map of building-block FNO, the largest error
source comes from the exterior boundaries of the simulation domain
where the prescribed Dirichlet BC changes too fast. Quantitatively,
the relative error averaged across all the fine nodes is less than
10\%, which is reasonably well for engineering applications.

\begin{figure}[H]
\centering \includegraphics[width=0.95\textwidth]{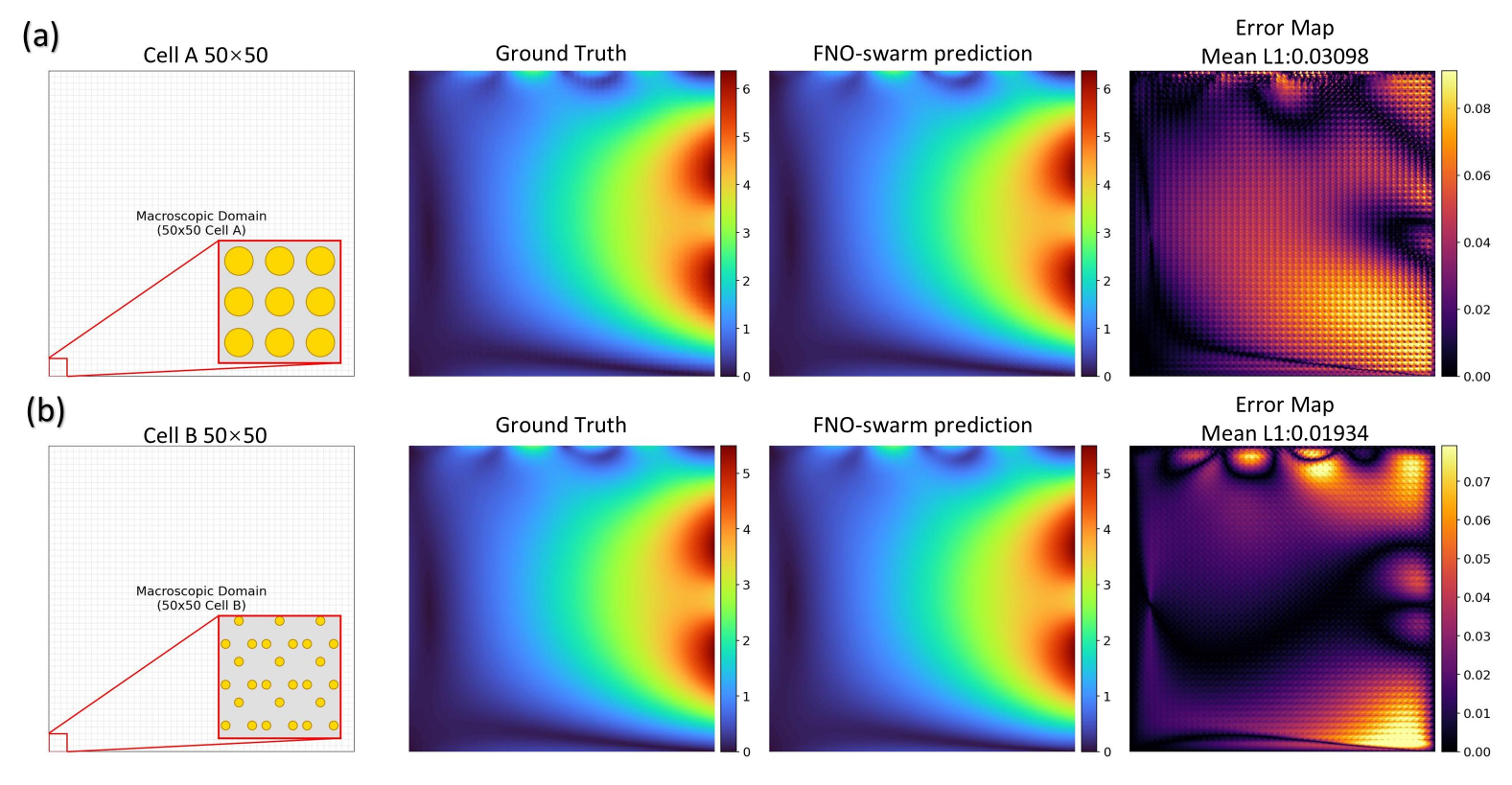}\caption{Test cases for the composite domains consisting of $50\times50$ unit
cell A and pure unit cell B. The second column shows the ground truths
of displacement magnitude obtained by fine-mesh FEM, while the third
column shows the corresponding results obtained by FNO-swarm method.
The last column shows the $L_{1}$ error map of the FNO-swarm results.}
\end{figure}

\subsubsection{Composites of mixed microstructure features}

The above validation cases demonstrate that the FNO-swarm method can
easily simulate the mechanical behavior of composites containing only
one microstructure feature. The real power of FNO-swarm method resides
in that it is designed to deal with arbitrarily complex arrangement
of the typical microstructure features. Fig.16 demonstrates a composite
domain formed by tiling unit cell A and B in alternating order, resulting
in a highly heterogeneous composite microstructure. The construction
of FNO swarm for this domain is similar as that in the previous cases,
except that the swarm now contains two type of FNOs corresponding
to unit cell A and B, respectively. Again, we employ both FNO-swarm
method and FEM to determine its mechanical responses under Dirichlet
BCs of $A=0.2$ and $B=1.0$. By comparing the calculated displacement
fields from FEM and from FNO-swarm method, we can see that the overall
patterns of the two displacement fields are in perfect alignment,
proving that FNO swarm is sufficiently flexible in dealing with such
heterogeneous mixture of microstructure features.

According to the error map, the average of point-by-point discrepancy
is estimated to be 0.0059, while the largest error is about 0.018,
which again is reasonably well for engineering applications. Moreover,
by analyzing the error distribution, we can see that the largest error
source now comes from the interior region instead of the exterior
boundaries. It suggests that the introduction of heterogeneous microstructure
has significantly increases the difficulty for collective inference
of the FNO swarm due to the mixing of different FNO models. Fortunately,
the error level is still acceptable to identify the regions of stress
concentration which is mainly determined by the general distribution
pattern of the displacement field, which under some occasions can
be of major concerns. More importantly, the interior error might be
effectively mitigated by further improving the prediction accuracy
of the building-block FNOs, which can be realized by extended training
or introducing more neural layers.

\begin{figure}[H]
\centering \includegraphics[width=0.95\textwidth]{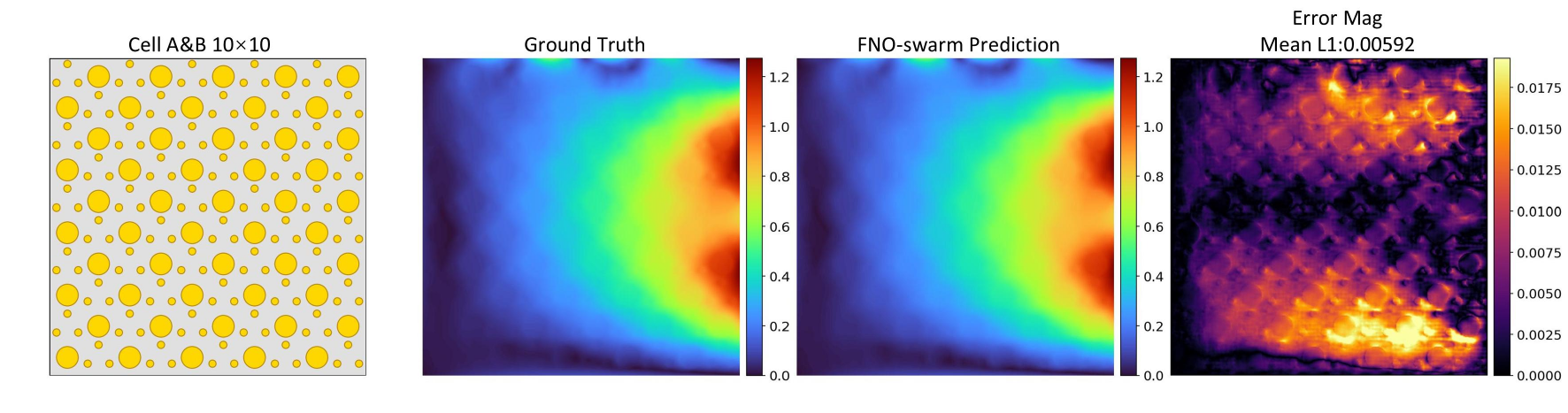} \caption{Test case for the composite domain consisting of $10\times10$ mixed
unit cell A and unit cell B.}
\end{figure}

This validation case implies that FNO swarm can deal with the random
mixture of arbitrary number of unit cells, as long as their mechanical
behaviors are accurately described by the the building-block FNOs
that have been trained in advance. Therefore, the FNO-swarm method
can be used to model the mechanical behavior of any composite domain,
as long as we extract all the typical microstructural features and
build FNOs for them. This characteristics significantly enhances its
flexibility in practical applications of simulating arbitrarily complex
composite microstructure. Such versatility makes it resemble a genuine
solution algorithm rather than a mere end-to-end mapping model, which
is particularly advantageous for the development of a general-purpose
numerical solver that can be applied across a wide range of simulation
tasks, just like classical FEM.

\subsubsection{Composites of exotic domain geometries}

For a general-purpose solution algorithm, it needs to be able to deal
with arbitrarily complex domain geometry. Since the typical unit cells
are defined as regular squares, the capability of FNO swarm in handling
extremely subtle geometries is intrinsically limited. However, we
can still come up with exotically shaped simulation domains made of
the regular squares, and show that the FNO swarm is sufficiently flexible
in dealing with these scenarios. For demonstration, we randomly generate
two different types exotic simulation domains, i.e., the composites
with internal porosity and L-shaped geometry.

First, let us investigate the performance of FNO-swarm method on porous
composite. We build the composite domain by first assembling $5\times5$
unit cell A, then randomly removing three unit cells into pores. Pure
Dirichlet boundary conditions are applied, with the left and bottom
edges fully clamped. The right edge is subjected to the prescribed
displacement $u=12.0sin(\pi s),v=0$, while the top edge is prescribed
with $u=0,v=15.0sin(2\pi s)$. When constructing the FNO swarms, we
only integrate those occupied cells and omit the pores. Afterward,
the Schwarz iteration is employed to achieve convergence and obtain
the mechanical response as in previous cases. Again, FEM calculation
result is used to obtain ground truths. Fig.17 compares the FNO-swarm
results of displacement field and equivalent strain field with the
corresponding FEM results. We can see that the two sets of results
are of the same overall patterns. To evaluate their quantitative discrepancies,
The error maps are calculated for both the displacement and strain
fields, according to which the average $L_{1}$ error of displacement
field is 0.254 and that of the equivalent strain is approximately
0.012. Moreover, according to the error maps,
the major error sources include the exterior boundaries where Dirichlet
BCs are applied and the corners of the pores. This is understandable,
since the the largest magnitude of the imposed BCs is much larger
than the average BC magnitude of the training set for the building-block
FNOs, which is expected to lead to degraded inference results. However,
the correct overall patterns of displacement and strain still prove
the merit of FNO-swarm method.

\begin{figure}[H]
\centering \includegraphics[width=0.95\textwidth]{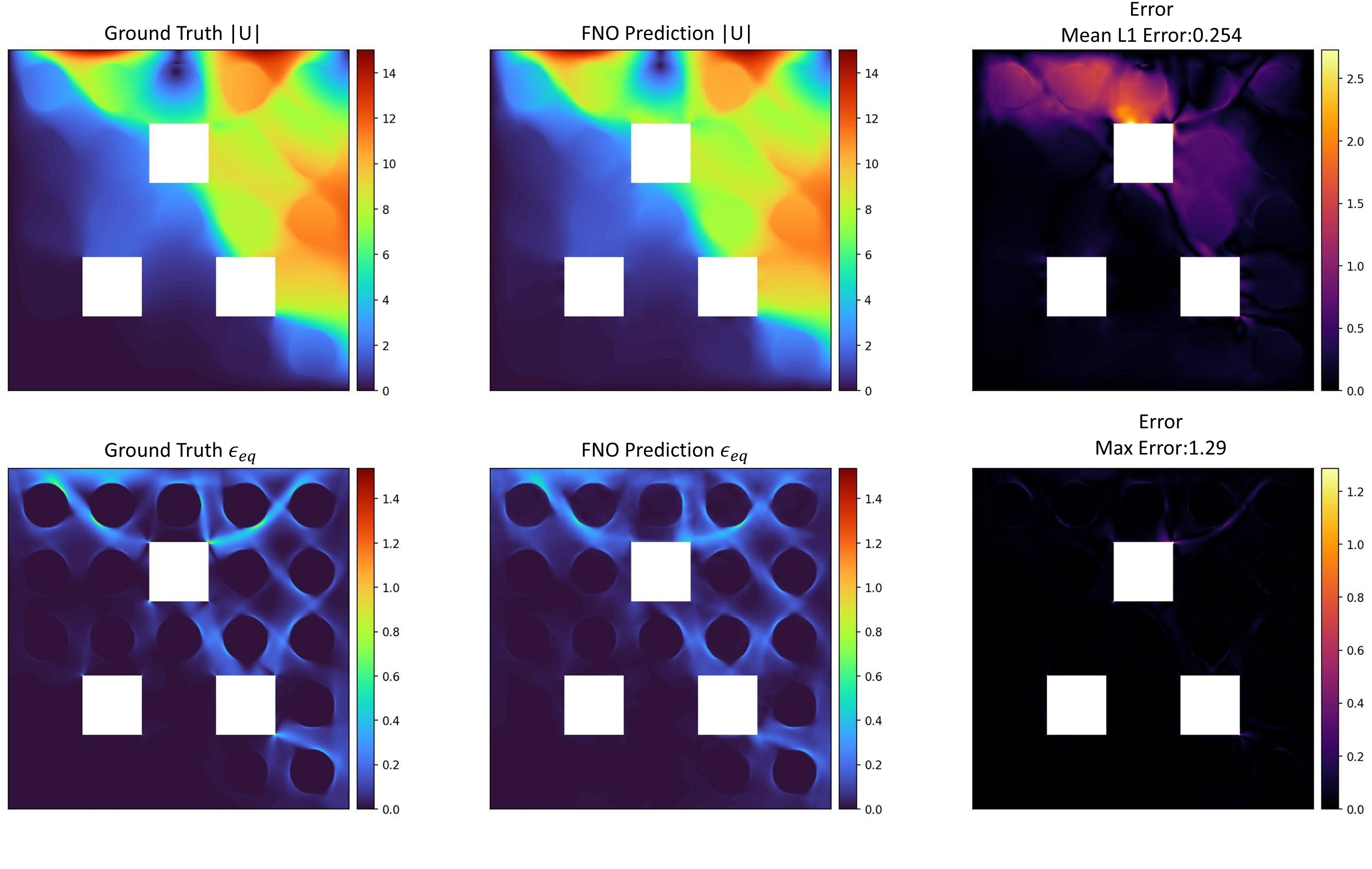}
\caption{Test case for the composite domain containing internal pores. The
simulation domain is constructed by first preparing an intact domain
of $5\times5$ unit cell A, then carving pores out of the domain.
The first column shows the FEM results of displacement and strain
fields, the second column is the FNO-swarm predictions, and the third
column is the error maps of the predictions.}
\end{figure}

Another example for irregular geometry is the L-shaped composite domain.
This domain consists of both unit cell A and B, as demonstrated in
Fig.18(a). The Dirichlet BCs is imposed on all the exterior edges
as illustrated in Fig.18(b). Fig.18(c)-(d) compares the displacement
field obtained by FEM and that predicted by the FNO-swarm method.
As we can see, the FNO-swarm prediction agrees well with the ground
truth in the entire domain. Quantitatively, the average error of displacement
across the entire domain is approximately $8.6\times10^{-4}$ while
the largest error is about $3.5\times10^{-3}$. The error map suggests
that the largest error occurs at reinforcement-matrix interface, which
is mainly due to the prediction error of the unit-cell FNOs. Again,
this error source can be expected to be mitigated by more thorough
training for the building-block FNOs.

\begin{figure}[H]
\centering \includegraphics[width=0.95\textwidth]{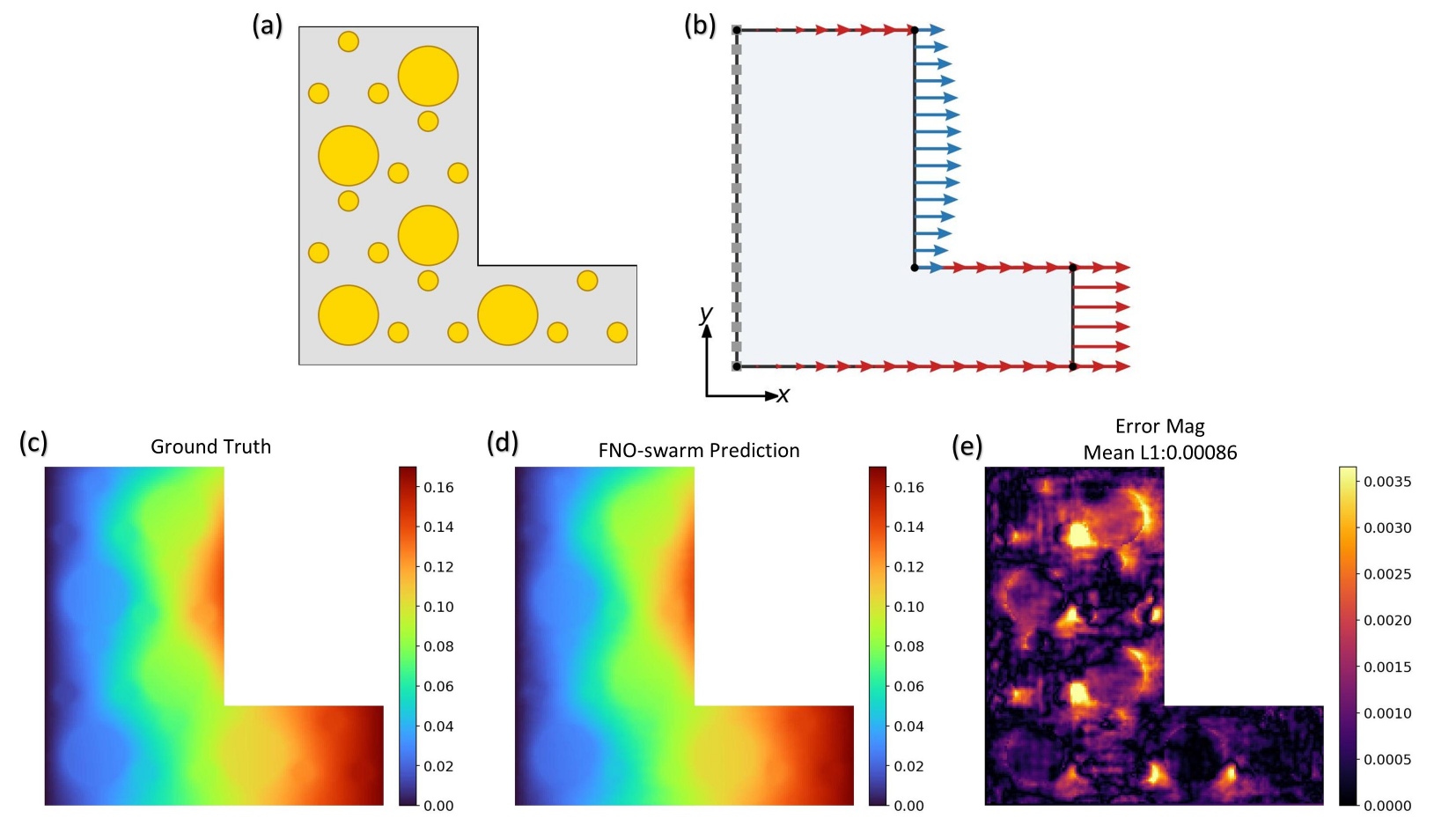}
\caption{Test case for the L-Shaped simulation domain consisting of mixed unit
cell A and B. (a)microstructure and domain geometry. (b)illustration
of the Dirichlet BCs. (c)the ground truth of displacement magnitude
field by FEM. (d) the predicted displacement magnitude field by FNO-swarm
method. (e) the error map of the FNO-swarm prediction.}
\end{figure}

These two examples demonstrate that the solving capability of the
proposed FNO-swarm method is insensitive not only to the configuration
of microstructural features but also to the geometry of the simulation
domain. This advantage is attributed to the "bottom-up" methodology
by assembling building-block FNOs flexibly into swarm based the microstructural
configuration and geometrical shape of composite domain. During the
assembly, the microstructural configuration and geometry shape are
dealt with by the non-computational linkage of the FNO models, while
the real computation has already been encoded in the FNOs of the unit
cells at offline training stage. As long as the typical microstructure
features remain the same, the FNO-swarm method can be employed for
arbitrary microstructural configuration and domain geometry.

\subsubsection{Composite with a dual-property dispersed phase}

It is well-known that the effective mechanical property of particle
reinforced composite material is closely related to the number density
and size distribution of the reinforcement particles. Therefore, we
can follow the similar idea as in manufacturing dual-property structural
component by maintaining fine sized and coarse sized grains at different
locations to satisfy different mechanical requirements for the corresponding
areas. Since we have designed two microstructure features captured
by unit cell A and B, representing the coarsely dispersed and finely
dispersed reinforcement phase, we can easily generate a dual-property
composite configuration containing separated A-rich region and B-rich
region. Such composites are especially suitable for large-scale structural
components, such as turbine disks, in which the central and peripheral
regions demand different mechanical properties to satisfy their respective
service requirements. For example, we usually require the bore region
exhibit superior strength and fatigue resistance while the rim region
provide enhanced creep resistance and high-temperature capability.
This tailored design improves engine efficiency, reduces weight, extends
service life, and enhances overall structural reliability.

The following example demonstrate the capability of the proposed FNO-swarm
method in dealing with dual-property composite of practical significance.
Specifically, the composite is constructed in the following procedure.
First of all, we create a square composite domain to contain $10\times10$
unit cells, whose rows are indexed by $r=0,1,\ldots,9$. For each
unit cell in row $r$, we first generate a random integer $X$ uniformly
from $[0,9]$, it is set to cell A if $X>r$ and B otherwise. This
stochastic assignment produces a gradual transition in microstructure
composition across the domain. The Dirichlet BCs are then imposed
using the same construction procedure as in Section 3.1, with $A=0.2$
and $B=1.0$. Both FNO-swarm method and FEM are employed to calculate
the equilibrium mechanical response of the composite domain. Fig.19
compares the overall pattern of the two results. As can be seen, the
FNO swarm has correctly captured the displacement distribution within
the domain. Moreover, quantitative analysis of the error map shows
that FNO swarm produces high-quality prediction at most of the regions.
The regions with relatively high error level are not the exterior
boundaries but the interior area where unit cell A and B are randomly
mixed. This error source has been discussed previously. The discrepancy
is most likely attributed to insufficient training of the building-block
FNOs and can be further mitigated through additional training or enhanced
model capacity. Consequently, the local error does not indicate an
inherent deficiency of the FNO-swarm architecture; instead, it underscores
the importance of developing more accurate and robust unit-cell surrogate
models.

\begin{figure}[H]
\centering \includegraphics[width=0.95\textwidth]{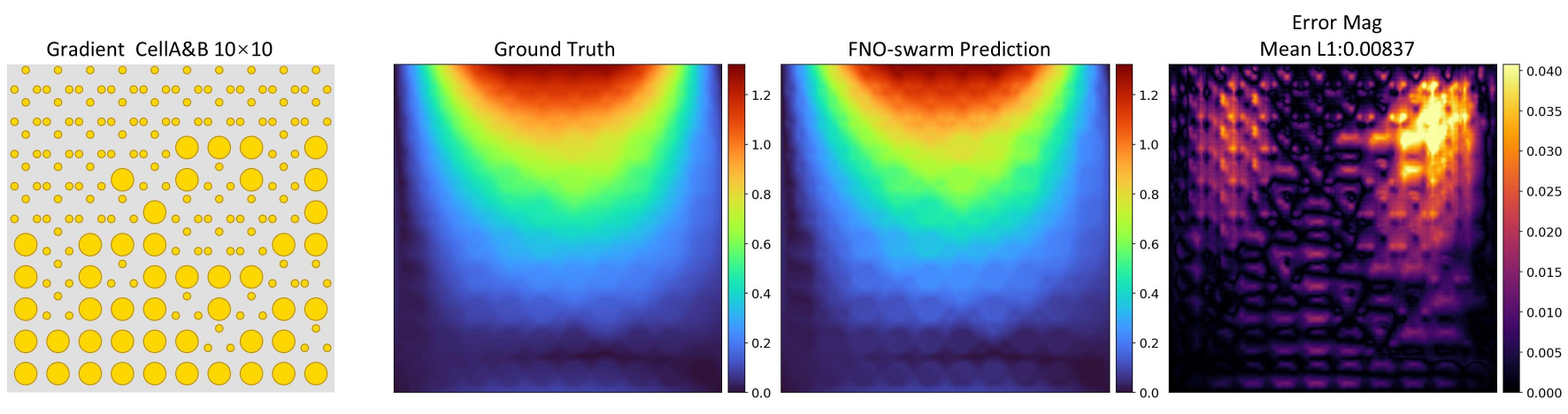}
\caption{Random Gradient Assembly}
\end{figure}

To rigorously assess the scalability and applicability of the proposed
FNO-swarm framework to extreme-scale multiscale problems, four ultra
large dual-property composite domains with complex shaped internal
cavities are investigated. The linear dimensions of each domain is
of $1000\times1000$ unit-cell width. The computational domains are
constructed by introducing cutouts of various topologies, including
elementary geometric shapes, alphabetic characters, and other abstract
patterns. The total number of sampling point within each domain is
$1.60\times10^{9}$, $2.05\times10^{9}$, $2.13\times10^{9}$ and
$2.10\times10^{9}$. The same Dirichlet BCs are imposed on all domains.
Specifically, the left boundary is prescribed with displacement field
$u=0,v=8sin(\pi s)$, the right boundary with $u=15s^{2}+5,v=5-5cos(2\pi s)$,
the top boundary with $u=15s^{2}+5s,v=10sin(\pi s)$, and the bottom
boundary with $u=5s,v=-20s(1-s)$. The results demonstrate that, on
a computing node equipped with a Intel Xeon Gold 6530 processor and
four NVIDIA RTX 4090 GPUs (24 GB each), the FNO-swarm method required
only 3735 seconds, 4488 seconds, 4653 seconds, and 4466 seconds, respectively,
to obtain the full-field solutions. In contrast, directly solving
models of this scale using classical FEM would incur computational
and memory requirements far beyond the capabilities of the current
hardware platform.

\begin{figure}[H]
\centering \includegraphics[width=0.95\textwidth]{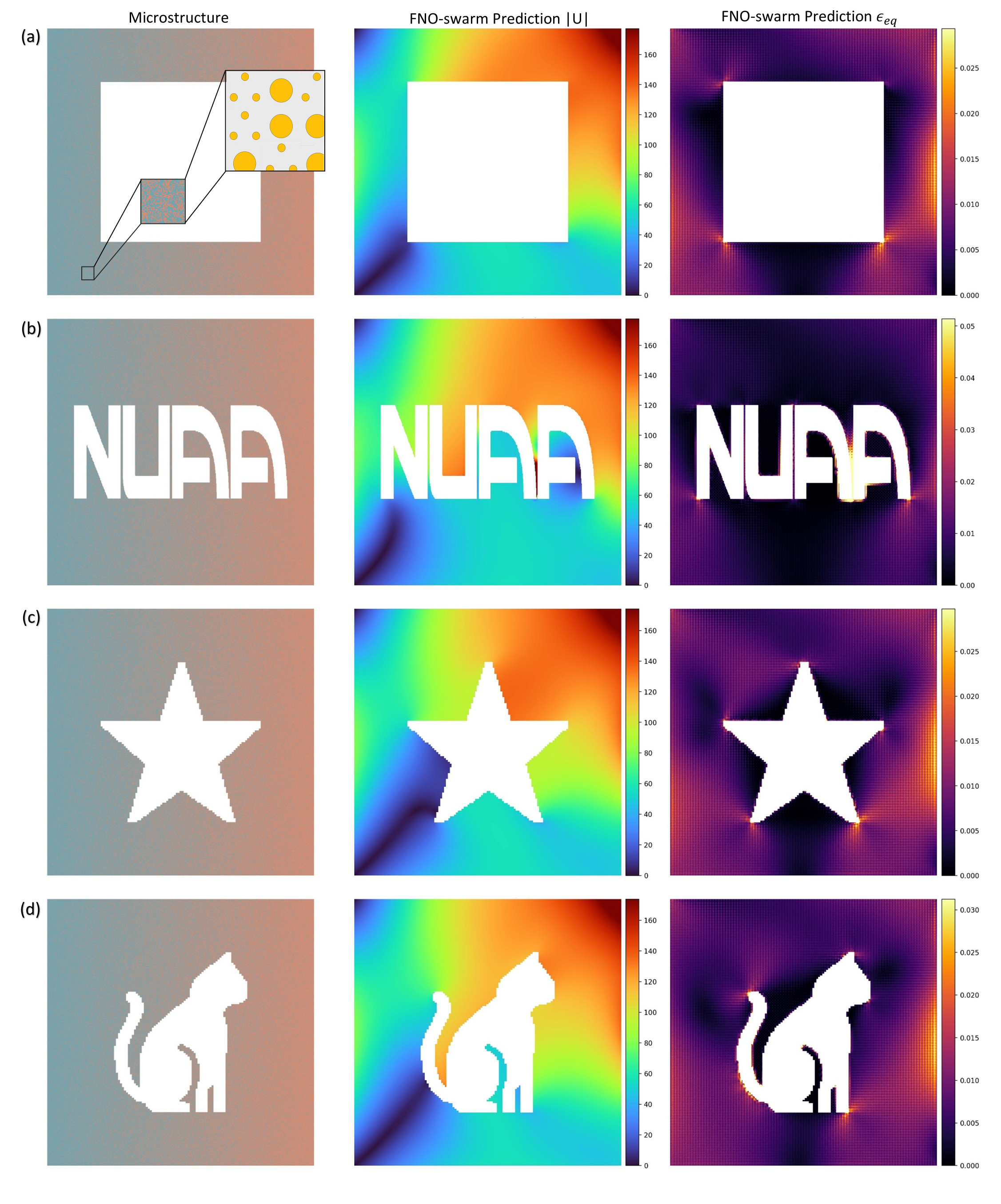} \caption{Ultra-large-scale unit-cell assembly with different internal cutouts.
The first column shows the graded microstructural distribution, where
red and blue regions denote Type A and Type B unit cells, respectively.
The second column presents the displacement magnitude predicted by
the FNO-swarm method. The third column shows the von Mises equivalent
strain field predicted by the FNO-swarm method.}
\end{figure}

Although the FNO-swarm method is proposed for 2D composite domain,
it is also applicable to 3D thin-walled structures without any modification.
To demonstrate this, we create a cylindrical shell, as shown in Fig.21(a).
Fig.21(b) demonstrates the unfolded microstructure, from which we
can see that the cylindrical shell structure consists of mixed cell
A and B, similar as the composite domain shown in Fig.16. Dirichlet
BCs are applied on the top and bottom edge of the shell. Specifically,
the top edge is prescribed by displacement field with $u=6.5sin(\pi s)$
and $v=6.5cos(\pi s)$. Here $u$ and $v$ represent the tangential
and axial displacement components, respectively, and $s$ represents
the arc length of the top edge. Similarly, the bottom edge is prescribed
by displacement with $u=-6.5sin(\pi s)$ and $v=-6.5cos(\pi s)$.
Fig.21(c)-(e) demonstrates the FEM result, the FNO-swarm result, as
well as the error map of these two. As can be seen, the FNO-swarm
prediction perfectly aligns well with the FEM ground truth except
at the end regions where BCs are imposed. It is worth noting that
this test case remains fundamentally a two-dimensional plane-stress
formulation and does not account for out-of-plane shear effects or
bending moments associated with classical shell theories. Feeding
the building-block FNOs with more specialized training datasets from
shell theories, the FNO-swarm method can also be used to accurately
capture the mechanical responses of real 3D thin-walled structures.

\begin{figure}[H]
\centering \includegraphics[width=0.95\textwidth]{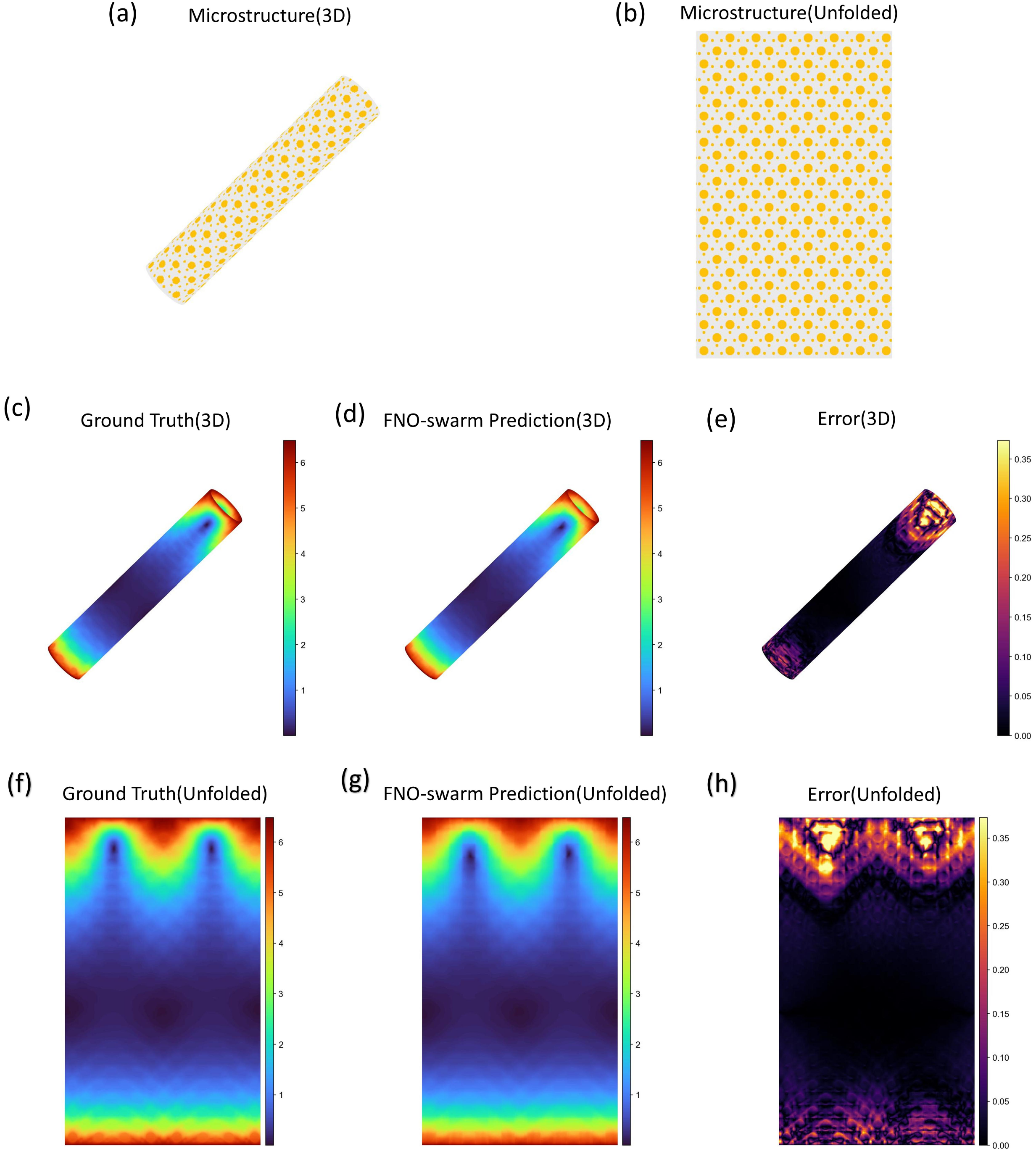} \caption{ (a)-(b)the microstructural configuration of the cylindrical shell
structure and its unfolded visualization. (c)-(e)comparison of the
FEM result and the FNO-swarm result of the displacement magnitude
field. (f)-(h)the unfolded visualizations of the FEM-calculated displacement,
the FNO-swarm-predicted displacement, and the error map. }
\end{figure}

\section{Analysis of physical fidelity and computational efficiency}

\subsection{Physical fidelity of FNO-swarm inference}

The black-box nature and poor generalization have been widely recognized
as the major drawback of ANN modeling in physical modeling. The former
usually leads to the absence of physical mechanism that is critical
for us to truely understand a physical mechanism, while the latter
often results in distorted results when the numerical condition goes
beyond the scope of training dataset. In the following example, we
demonstrate the performance of FNO-swarm method on a case where the
applied Dirichlet BCs significantly exceed the range of the training
data. We will not only compare its direct inference result of displacement
field, but also investigate the derived physical behavior of stress
concentration.

The composite contains $5\times5$ unit cell A. The Dirichlet BCs
are set as follows. The left and bottom boundaries are completely
constrained. The right boundary is prescribed with displacement field
of $u=12.0sin(\pi s),v=0$, while the top boundary is prescribed with
$u=0,v=15.0sin(2\pi s)$. Here the largest magnitude of the applied
displacement component is well beyond the scope of the training data
prepared for the FNO of unit cell A. Fig.22 compares the magnitude
of displacement field predicted by FNO-swarm method and that by FEM.
As can be seen, the overall patterns are in good alignment. The displacement
error map confirms that most of the regions have quite low discrepancy
except one region that shows slightly larger error. The overall agreement
between the FNO-swarm and FEM suggest that the FNO-swarm method can
reliably produce reasonable result even when the imposed BCs are not
covered by the training data. Moreover, the comparison between the
derived equivalent strain fields from FEM and FNO-swarm method shows
that, despite a few locations of relatively large discrepancy, the
two strain fields are generally in rather acceptable agreement. Both
of them demonstrate the same stress concentration distribution, suggesting
that the FNO swarm can correctly capture the underlying physical mechanism.
This case indicates that, even for the scenario far away from the
training dataset, the FNO-swarm can still produce physically reasonable
result. We believe that this generalization capability is mainly attributed
to the collective inference of FNOs within the swarm. The collective
inference requires the mutually linked FNOs behave consistently, which
helps to eliminate unphysical fluctuations and discontinuities.

\begin{figure}[H]
\centering \includegraphics[width=0.95\textwidth]{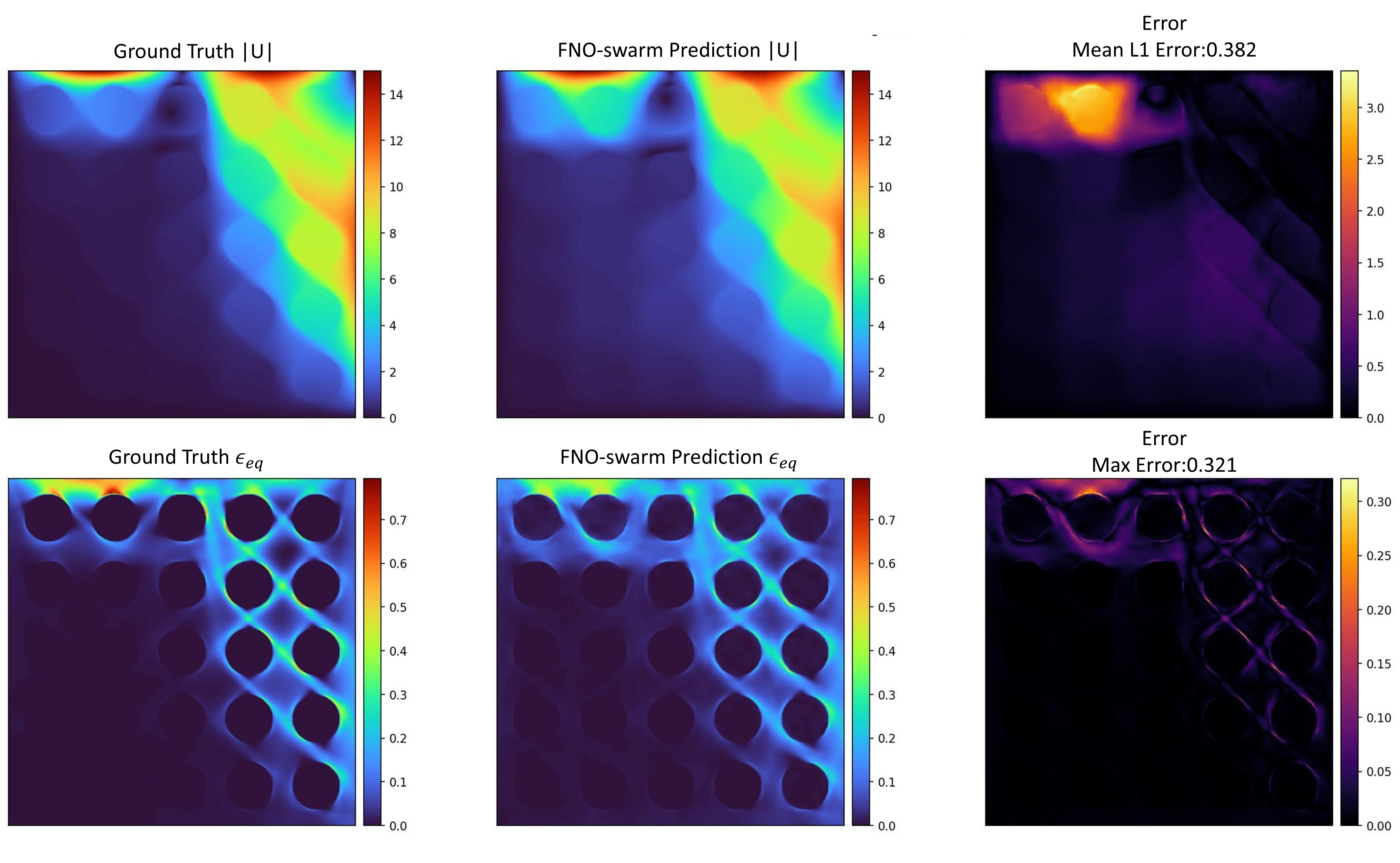}
\caption{Test for the out-of-scope case. The first column is the FEM results
of displacement magnitude and equivalent strain, the second column
is the corresponding FNO-swarm results, the third column is the $L_{1}$
error maps.}
\end{figure}

To investigate when and how the FNO-swarm calculation deviates from
the ground truth from FEM, we have tried to impose the BCs on the
composite domain progressively, and monitor the degradation of the
FNO-swarm result. Specifically, based on the above simulation problem,
we have performed a series of similar simulations by imposing 20\%,
50\%, 80\%, and 100\% of the original Dirichlet BCs. For each simulation,
we compare the the average strain and stress obtained by FNO-swarm
method and FEM. Specifically, after obtaining the displacement field,
we first calculate the node-wise strain components $\epsilon_{xx}$,
$\epsilon_{xy}$, $\epsilon_{xy}$, then obtain the node-wise stress
component $\sigma_{xx}$, $\sigma_{xy}$, $\sigma_{xy}$ by utilizing
the constitutive models of the Al matrix and SiC reinforcement. For
the strain and stress filed, we calculate the node-wise equivalent
strain $\epsilon_{eq}$ and von Mises stress $\sigma_{m}$. Finally,
we obtain the effective strain and stress by calculating the average
values of $\epsilon_{eq}$ $\sigma_{m}$ across the entire domain.

Fig.23 plots the average strain and stress for each simulation. As
can be seen, for small strain scenario, the effective strain and stress
resulted from FNO swam is in perfect agreement with those from FEM.
As the strain increases, the deviation gradually expands. Such deviation
is mainly originated from the degraded prediction accuracy of the
building-block FNOs of the unit cells, which is a direct consequence
of the strain magnitude going beyond the coverage of the training
dataset. The error source is expected to be effectively reduced by
expanding the training dataset. Despite the increasing deviation,
the primary evolution trend of the effective strain-stress relation
remains consistent, which again suggests that the FNO-swarm method
is robust when dealing with out-of-scope situations, rather than producing
nonsensical results.

\begin{figure}[H]
\centering \includegraphics[width=0.6\textwidth]{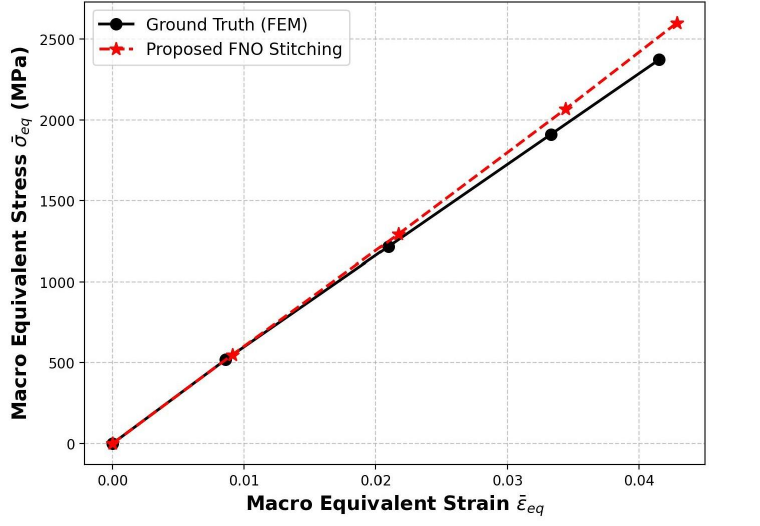}
\caption{Investigation of the degradation of FNO-swarm method as the boundary
conditions gradually exceed the coverage of the training dataset.}
\end{figure}

\subsection{Initialization dependence}

As introduced before, the FNO-swarm method requires initialization
for the edges of unit-cell subdomain. Specifically, we treat the entire
composite domain as a homogeneous single phase of Al matrix and generate
a coarse mesh. Then, We employ FEM to calculate the displacement field
on this coarse mesh and use bicubic interpolation to initialize the
input layers of the unit-cell FNOs. Such initialization scheme is
efficient as long as the the coarse mesh doesn't contain too many
elements. the simplest coarse mesh is regarding each unit-cell subdomain
as a giant quadrilateral element so that number of elements is just
the number of unit-cell subdomains. To determine the relation between
the inference accuracy of the FNO swarm and the size of initializing
coarse mesh, we have performed a series of calculations over a composite
domain with the same Dirichlet BCs and Schwarz iteration number, but
different mesh for initialization. The composite consists of $10\times10$
unit cell A. Denoting the element number along each dimension of the
coarse mesh by $N_{init}$, the calculations are initialized with
$N_{init}=0\sim50$. When $N_{init}=0$, the coarse-mesh FEM is not
defined, we employ bilinear Coons interpolation to initialize the
building-block FNOs. Coons interpolation, originally introduced by
Steven A. Coons, provides a means of generating a smooth interior
field that exactly satisfies prescribed boundary conditions \cite{1967Surfaces}.
It should be emphasized that the resulting field enforces only geometric
consistency and smoothness constraints, without necessarily satisfying
the underlying physical governing equations.

Fig.24(a) shows the relation between the prediction error level of
FNO-swarm method on the coarseness of the initialization mesh. As
can be seen, the prediction error doesn't further decrease once the
coarse mesh size $N_{init}$ reaches 10. This is the scenario when
the coarse mesh is constructed from the corner points of the subdomain
and each subdomain becomes a single quadrilateral element of the coarse
mesh. With this level of coarseness combined with the assumption of
single-phase simulation domain, the time cost of coarse-mesh FEM initialization
has virtually no negative effect on the computational efficiency of
the FNO-swarm method. Fig.24(b)-(d) demonstrates the calculated displacement
fields with different initialization coarseness, from which we can
see that the FNO-swarm result appears decent even when $N_{init}=0$.
These results suggest that FEM initialization can significantly help
the convergence of FNO swarm. Meanwhile, initialization with a coarse
mesh and single-phase assumption is sufficient to produce the desired
convergence.

\begin{figure}[H]
\centering \includegraphics[width=0.95\textwidth]{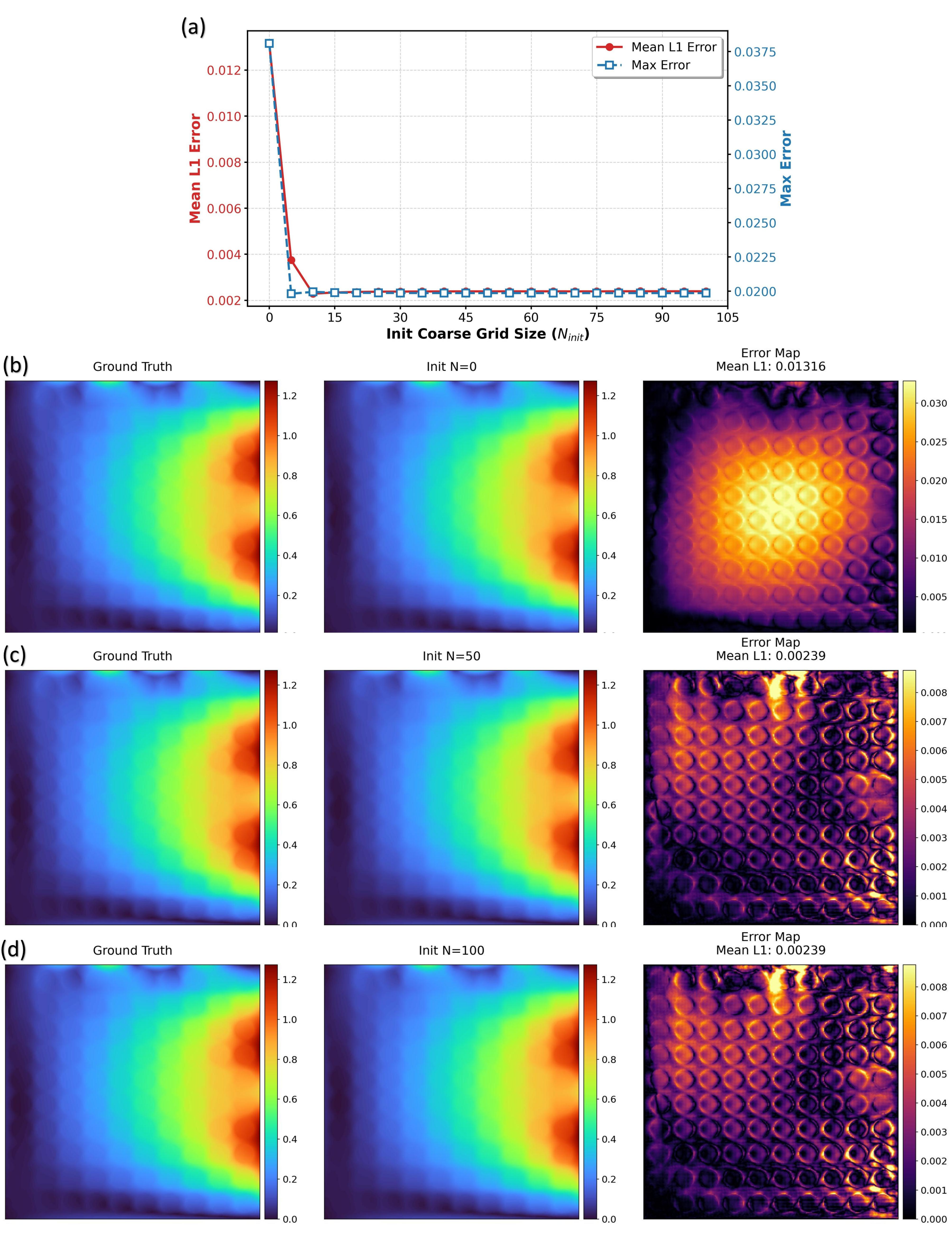} \caption{(a)evolution of FNO-swarm prediciton error with respect to the coarseness
of the initialization mesh. (b)-(d)The displacement fields predicted
by FNO-swarm method under different initialization mesh sizes.}
\end{figure}

\subsection{Noise resistance}

A major strength of neural network modeling is that it can take arbitrary
legitimate data source. Although we use FEM here to collect training
dataset for the unit-cell FNOs, it is equally feasible to train the
FNOs with experimental data. However, the data collected from experiments
unavoidably contains noise from measurement. For non-robust solving
method, the occurrence of noise can be lethal and completely destroy
the credibility of solution. To assess the noise resistance of the
proposed FNO-swarm method, we will superimpose a high-frequency fluctuation
onto the applied Dirichlet BCs, whose mathematical form is $N(x,y)=C\cdot sin(\omega x)\cdot sin(\omega y)$.
The simulation domain is a L-shaped composite slab purely consisting
of unit cell A. The Dirichlet BCs are set similarly as before, except
that a noise with $C=0.02$ is added. We have performed a series of
calculation with noise frequency $\omega\in[0,100]$. For each calculation,
we obtain the FNO-swarm result under the noise-affected BCs and compare
it with FEM result under the noise free BCs by evaluating the mean
and maximum discrepancies throughout the simulation domain.

Fig.25(a) demonstrates the error evolution against the noise frequency.
As can be seen, for the investigated spectrum, both the mean and maximum
errors maintain approximately the same level with some local fluctuations.
Therefore, the introduction of uncertainty in the BCs cannot lead
to catastrophic failure in the FNO-swarm results. Considering that
the training data collection never includes violently changing displacement
fields, the FNO-swarm method shows strong robustness in dealing with
noisy BCs. Fig.25(b)-(c) compares the displacement fields calculated
by FEM and FNO-swarm method under different noise frequencies. According
to the error map, we can see that the FNO-swarm inference under noise
polluted BCs is generally in good agreement with the ground truth
except at the exterior boundaries where the BC noise is enforced.
It clearly shows that the degradation led by the introduced noise
is constrained to the exterior edges while the interior region is
largely not affected. Such noise resistance mathematically stems from
the error attenuation characteristics of the Schwarz iteration. Physically,
it aligns with Saint-Venant's principle, where high-frequency perturbations
at boundaries rapidly dissipate as they propagate inward into the
domain. This again indicates the high physical fidelity of FNO-swarm
method. Therefore, the FNO-swarm method is rather tolerant to the
noisy BCs by producing legitimate general pattern regardless of introduced
BC perturbations. This is of great practical significance when the
true boundary conditions can not be accurately prescribed.

\begin{figure}[H]
\centering \includegraphics[width=0.95\textwidth]{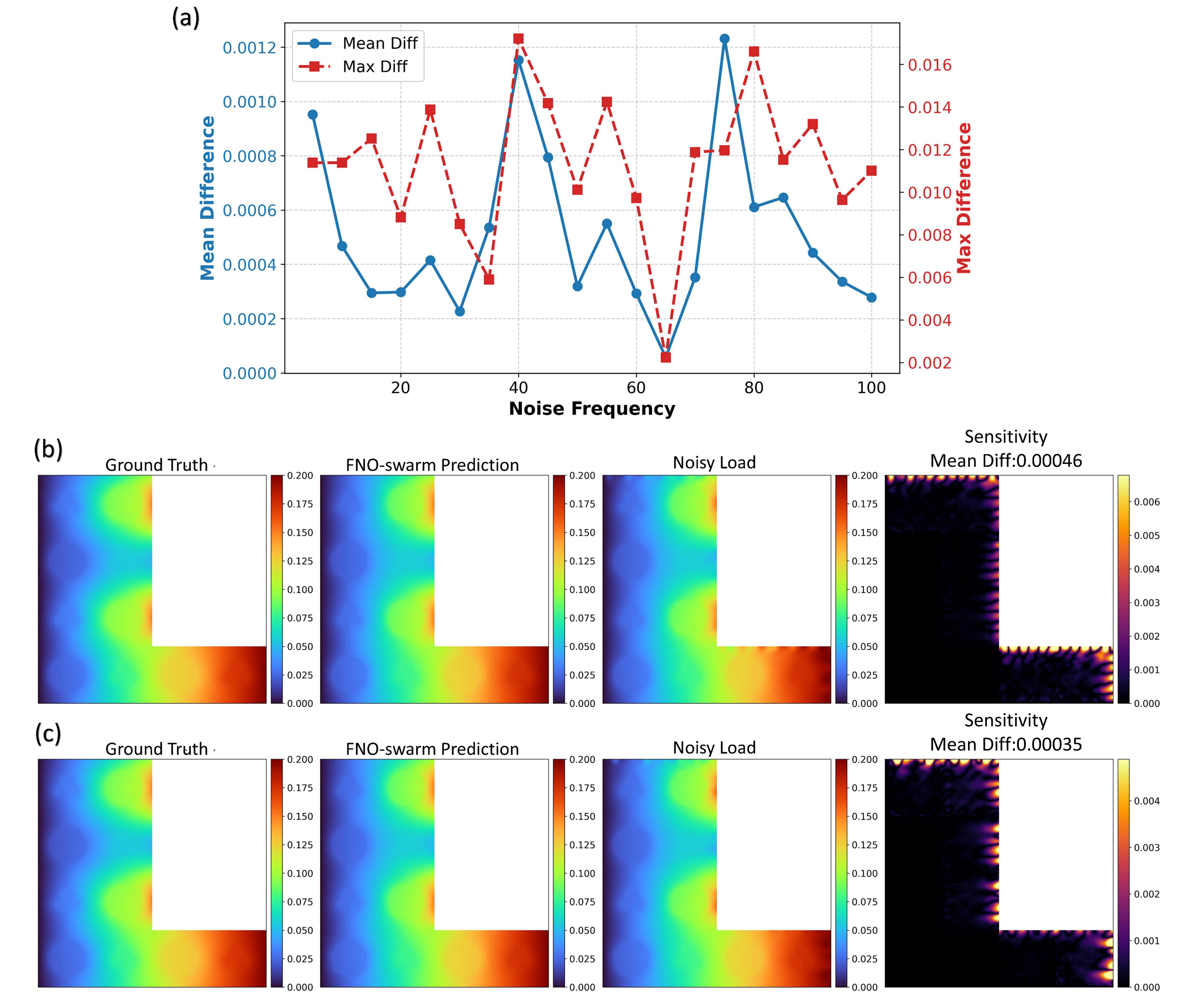}
\caption{(a)the average error and maximum error evolution with respect to the
noise frequency when the noise amplitude is fixed to $C=0.02$. (b)(c)the
FNO-swarm-predicted displacement field under noise frequency $\omega=50$
and $\omega=100$. The last column shows the difference between the
prediction results before and after the addition of noise.}
\end{figure}

\subsubsection{Convergence pattern and computational scaling}

We have theoretically analyzed that the error of the Schwarz iteration
is capped by prediction error of the building-block FNOs and the overlapping
width between subdomains. However, the theoretical conclusion is derived
under the assumption that number of iterations can be as long as needed.
In real applications, we don't have such luxury where efficiency is
an important consideration. To investigate the converging characteristics
of the FNO-swarm method, we randomly generate 1000 various Dirichlet
BCs and impose them on the composite configuration with $5\times5$
unit cell A. We keep record of the converging behavior for each case
by monitoring the evolution of the solution vector. Specifically,
at $k_{th}$ iteration, we calculate the relative difference between
current solution and the solution from previous step, i.e., $|u^{(k)}-u^{(k-1)}|/|u^{(k-1)}|$.

Fig.26 demonstrates the convergence pattern of the 1000 cases in a
single plot on semi-log scale. For all cases, the relative variation
of the solution vector drops drastically below one percent in the
first dozens of iterations. Afterward, the relative variation of the
solution vector decreases exponentially with increasing iterations.
This convergence behavior has two implications. First, we can expect
very efficient collective coordination of the subdomain FNOs once
the Schwarz iteration starts. Second, when the inter-subdomain inconsistency
decreases below some threshold, a long tail effect occurs when the
error reduction is relative slow but the trend of convergence does
not change. Which means that after sufficiently long iterations, the
desired convergence can be achieved as described by the theoretical
analysis. Fortunately, while inspecting the evolution of the L1 loss,
we find that even under the worst converging behavior (the upper envelope
in Fig.26) the loss function still goes down to approximately $10^{-}6$
after 150 iterations. Therefore, the proposed FNO-swarm method is
of excellent convergence behavior, a necessary merit for a general-purpose
solving algorithm.

\begin{figure}[H]
\centering \includegraphics[width=0.95\textwidth]{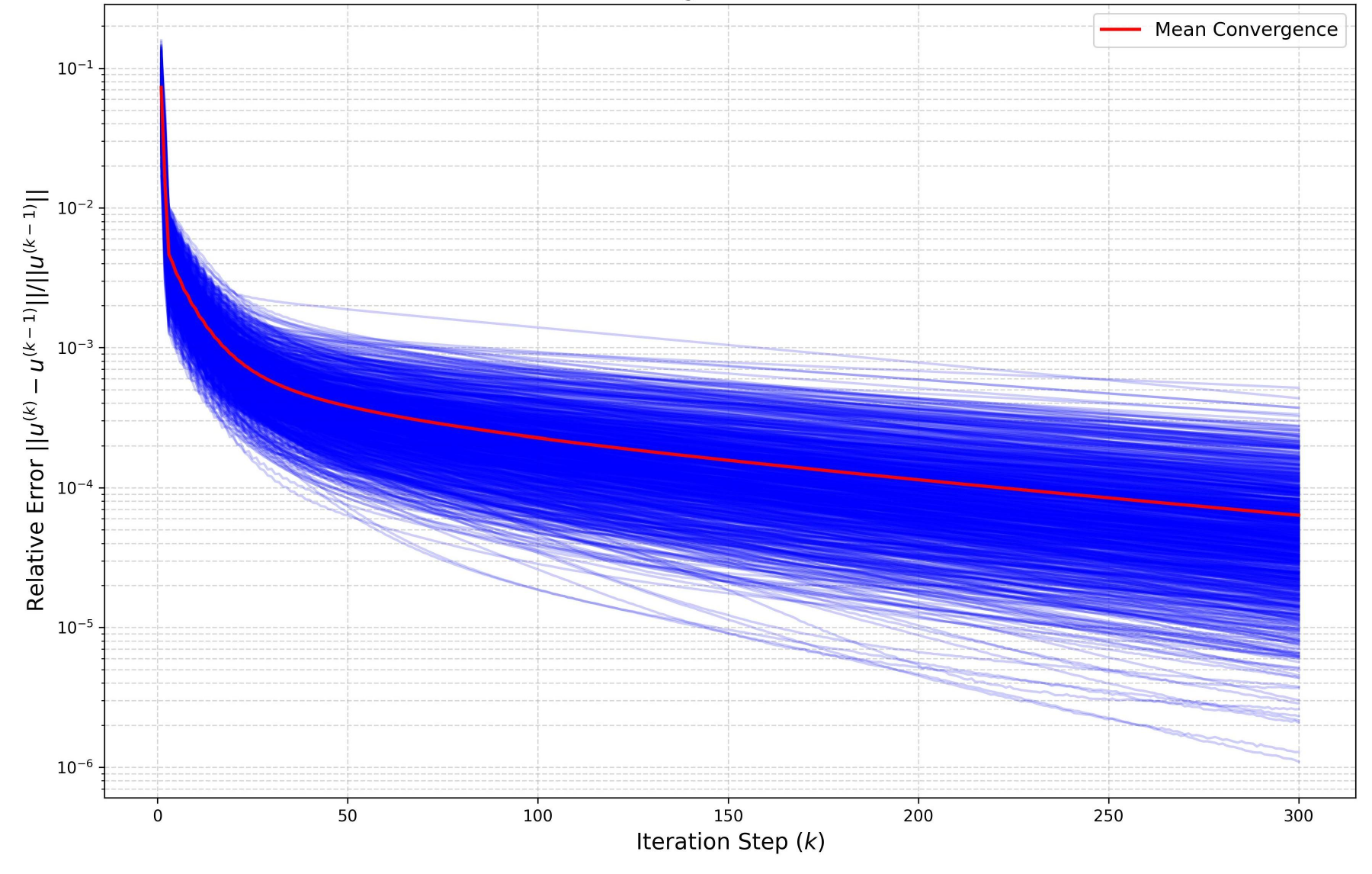}
\caption{Convergence curves for randomly generated boundary conditions}
\end{figure}

For many classical numerical algorithms, the accelerated growth of
computational cost with respect to problem size are the major hurdles
for them to deal with large-size simulation tasks. Typical examples
include molecular dynamics, phase field modeling, FEM, etc. Designing
advanced algorithms with slower cost growth has been a long-lasting
goal in computational mechanics. To investigate the computational
scaling characteristics of the proposed FNO-swarm method, we have
prepared a series of composite simulation domains with different sizes.
Specifically, all the composites are purely made of unit cell A whose
side length ranges from 5 cell widths to 50 cell widths, and the same
pattern of Dirichlet BCs are applied to each domain. However, since
the convergence rate of FNO swarm is dependent on the material nonlinearity,
we linearly scale the magnitude of the applied BCs with respect to
the domain size to achieve an approximately 3\% average strain for
all the cases so that the computational costs are compared against
the same baseline. For reference, the FEM calculation is also carried
out for each simulation case, during which each subdomain is meshed
into approximately $10^{4}$ nodal points. For FEM calculations, memory
consumption is also a critical cost, so we have also recorded the
memory usage for each calculation. Here, the FEM solver is implemented
in C++ and executed on a CPU, while the FNO-swarm framework is implemented
in Python and executed on a GPU.

Fig.27 demonstrates the problem-size dependence of time cost (on single
processor), memory consumption, and accuracy of the FNO-swarm , compared
with those of FEM. We can easily recognize that the computational
time of both FNO-swarm method and FEM increases exponentially with
problem size, but the former is generally $1\sim2$ magnitudes smaller
than FEM. Moreover, the growth rate of FNO-swarm method is also much
smaller than FEM. Similar trend can be found in the memory usage comparison,
where FEM is generally much more memory-demanding than FNO-swarm method.
This is because FEM requires the construction and global stiffness
matrix and heavy linear algebra operations, e.g., matrix decomposition,
inversion, which usually consumes large amount of memory storage.
While for the FNO-swarm method, the storage requirement of FNO models
are negligible, and the major consumer of memory is the input and
output vector, which is one dimension less than the stiffness matrix
of FEM. Thus, memory efficiency is one of prominent features of FNO-swarm
method. Most importantly, the superiority of FNO-swarm method in computational
and memory advantage is attained without sacrificing its solution
accuracy. We can see that the average error of the FNO-swarm result
increases linearly with the problem size. Remembering that we have
increased the magnitude of the applied Dirichlet BCs linearly with
respect to the problem size in order to maintain approximately constant
strain level, the linearly increasing absolute error means that the
relative error is nearly fixed for all the simulation cases, regardless
of problem size. Although the comparisons should be interpreted in
light of the different software and hardware environments, the scaling
trends in computational time and memory usage and approximately fixed
relative error still confirm that the proposed FNO-swarm method has
the potential to serve as an efficient and reliable computational
tool with modest hardware requirements.

\begin{figure}[H]
\centering \includegraphics[width=1\textwidth]{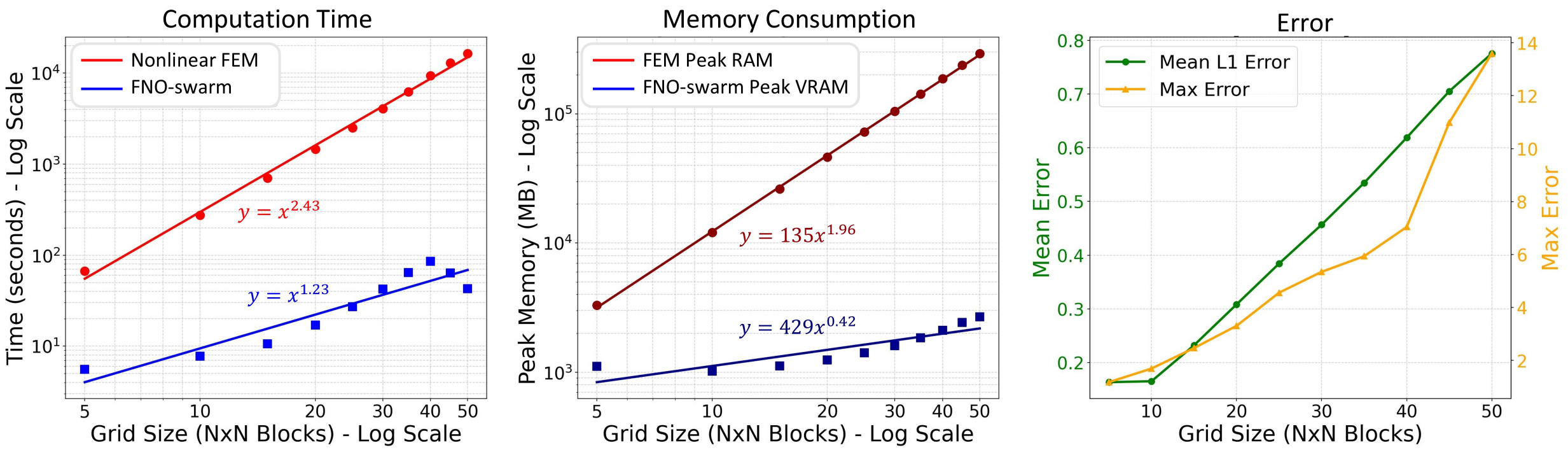} \caption{Investigation of computational time cost, memory consumption, and
result accuracy of FNO-swarm method with respect to different problem
sizes.}
\end{figure}

\section{Discussion}

The development of neural network architectures specifically tailored
to computational mechanics has attracted considerable research attention
in recent years\cite{0Physics,WU2026118799,Herrmann2024DeepLI,2017The,MASI2021104277}.
The goal is to achieve flexibility in dealing with irregular simulation
domain, arbitrary BCs and complex internal microstructure efficiently
and reliably by tailoring the neural network modeling strategy. The
proposed FNO-swarm method leverages domain decomposition and Schwarz
iteration to achieve this goal through three key design components.
First, instead of directly mapping from microstructure to mechanical
response, it segments the overall microstructure into a couple of
typical features represented by unit cells and use level set method
to convert discrete microstructure into nearly a continuous field,
which significantly reduce the difficulty of FNO training. Second,
it utilizes coarse-mesh FEM calculation result as initialization,
which serves as initial guidance for the Schwarz iteration to quickly
find the evolution direction with negligible computational cost. Third,
based on the coarse-mesh FEM initialization, the FNO and FNO-swarm
only need to the learn the residual between the true solution and
the initial guess. Since the residual mainly contains fluctuating
variations, which can be efficiently captured in the spectral domain.
These design features renders the proposed FNO-swarm method an efficient
and accurate general-purpose solving algorithm.

However, there are still some caveats needs to be clarified. Since
the FNO swarm is made of building-block FNOs for unit cells, how to
extract the specific unit cells is the key to employ the FNO-swarm
method. However, here we simply manually determine the unit cells
since the microstructural configuration of particle reinforcement
composite is relatively clean. For more complex microstructure, it
is necessary to design automatic approach to fulfill this task. Another
issue is with the geometry of unit cell. Different from FEM which
defines a plethora of elements with various shapes, the FNO-swarm
method currently only take regular shaped unit cells, which may not
be sufficient to describe extremely complex shaped structural components.
Finally, despite the its decent accuracy in reproducing the displacement
field, the derived strain field still shows quantitatively large error
at exterior boundaries where strongly fluctuating BCs are imposed,
and the absolute level of strain error clearly increases with problem
size. Fig.28 compares the displacement and strain fields obtained
by FNO-swarm method and that by FEM for both small-sized and large-sized
simulation domains. As can be seen, even though the overall patterns
of the strain field is correctly captured by FNO swarm, the maximum
strain error can be of the same order as the strain value itself for
very large problem.Possible strategies to address this issue include
enhancing the prediction accuracy of the building-block FNOs and incorporating
loss terms designed specifically for the displacement gradient.

\begin{figure}[H]
\centering \includegraphics[width=0.95\textwidth]{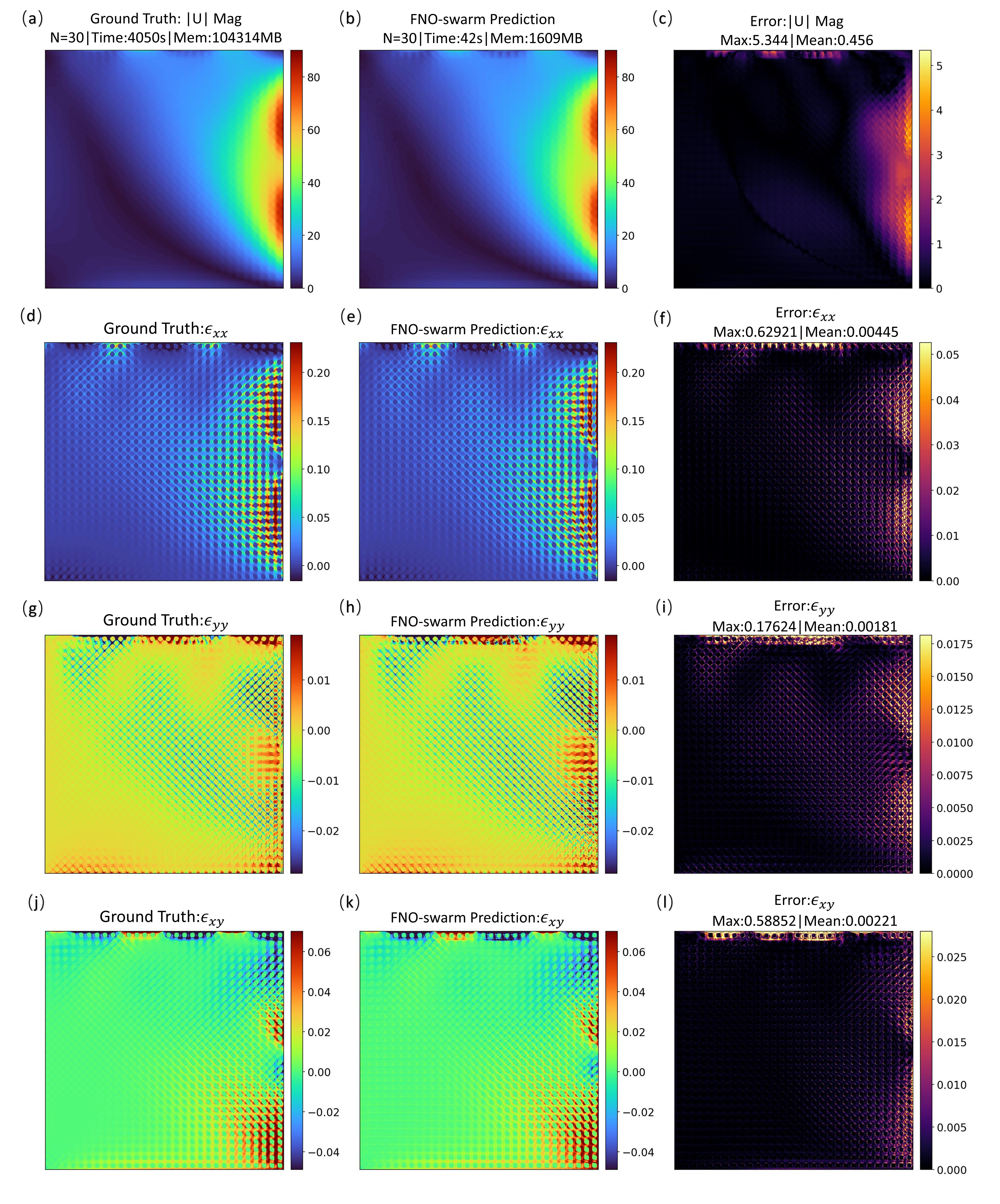}\caption{The displacement field and strain components predicted by FNO-swarm
method for large simulation domain containing $30\times30$ unit cell
A. (a)-(c)shows the ground truth, FNO-swarm prediciton and error map
of displacement magnitude. (d)-(f)shows the ground truth, prediction
and error map of strain component $\epsilon_{xx}$. (g)-(i)shows those
of strain component $\epsilon_{yy}$. (j)-(l)shows those of strain
component $\epsilon_{xy}$.}
\end{figure}

\section{Conclusion}

Dedicated to designing a fully scalable general-purpose algorithm
for efficient simulation of composite media containing complex microstructure,
we have introduced a FNO-based machine learning strategy which leverages
domain decomposition, residual learning and collective intelligence
of a swarm of FNO models. It begins by representing the characteristic
microstructural constituents of a composite material using building-block
FNOs, which are then flexibly organized into a hierarchical FNO swarm
according to the microstructural configuration of the target simulation domain. The
Schwarz iteration method is subsequently employed to coordinate the
building-block FNOs, enabling them to collectively synchronize to a global
solution of the mechanical responses. Extensive simulations on
the Al--SiC composite system demonstrate that the proposed FNO-swarm
method achieves both high computational efficiency and excellent accuracy.
Under identical conditions, it reduces computational time and memory
consumption by approximately two orders of magnitude compared with
the classical FEM. Moreover, its weak dependence on initialization
and strong resilience to high-frequency noise suggest that it can
be potentially used as a general-purpose numerical solver alongside
conventional numerical methods. Nevertheless, the proposed FNO-swarm method
still has several limitations, including compromised accuracy in derived
quantities such as strain fields and the restriction on the geometric
shape of unit cells. Addressing these issues will be a primary focus
of future work.

\section{Data availability}

The data and codes involved in this work are available upon reasonable
request.

\section{Acknowledgement}

This work is supported by the National Natural Science Foundation
of China (No. 12302178) and the Natural Science Foundation of Jiangsu
Province (No. BK20230897). The authors acknowledge Beijng PARATERA Tech CO.,Ltd. 
for providing HPC resources that have contributed to the research results reported 
within this paper. URL: https://paratera.com/.


 \bibliographystyle{unsrt}
\bibliography{reference}

\end{document}